\documentclass[epj,nopacs]{svjour}
\usepackage{graphicx}

\begin{document}
\title{
The role of strange sea quarks in\\
chiral extrapolations on the lattice}
\author{
S\'ebastien Descotes-Genon
}                     
\institute{Laboratoire de Physique Th\'eorique\thanks{LPT is an Unit\'e Mixte de Recherche du CNRS et
de l'Universit\'e Paris-Sud 11 (UMR 8627).}, 
Universit\'e Paris-Sud 11, 91405 Orsay Cedex, France}
\date{}
%
\abstract{Since the strange quark has a light mass of order $O(\Lambda_{QCD})$,
fluctuations of sea $s\bar{s}$ pairs may play a special role in the low-energy
dynamics of QCD by inducing significantly different patterns
of chiral symmetry breaking in the chiral limits $N_f=2$ ($m_u=m_d=0$, $m_s$
physical) and $N_f=3$ ($m_u=m_d=m_s=0$). This effect of vacuum fluctuations
of $s\bar{s}$ pairs
is related to the violation of the Zweig rule in the scalar sector, described
through the two $O(p^4)$ low-energy constants $L_4$ and $L_6$ of 
the three-flavour strong chiral lagrangian. In the case of significant
vacuum fluctuations, three-flavour chiral expansions might exhibit
a numerical competition between leading- and next-to-leading-order
terms according to the chiral counting, and chiral extrapolations should 
be handled with a special care. We investigate the impact of 
the fluctuations of $s\bar{s}$ pairs on chiral extrapolations in the 
case of lattice simulations with three dynamical flavours
in the isospin limit. Information 
on the size of the vacuum fluctuations can
be obtained from the dependence of the masses and decay
constants of pions and kaons on the light quark masses.
Even in the case of large
fluctuations, corrections due to the finite size
of spatial dimensions can be kept under control for large enough boxes
($L\sim 2.5$ fm).
}
\maketitle
In order to achieve a better understanding of
nonperturbative features of the
strong interaction, it is interesting to recall the particular
mass hierarchy followed by the light quarks:
\begin{equation}
m_u\sim m_d \ll m_s \sim \Lambda_{QCD} \ll \Lambda_H\,,
\end{equation}
where $\Lambda_{QCD}$ is the characteristic scale describing the
running of the QCD effective coupling and $\Lambda_H\sim 1$ GeV the mass
scale of the bound states not protected by chiral symmetry. Therefore,
the strange quark may play a special role in the low-energy dynamics of
QCD: 

\emph{i)} it is light enough to allow for a combined expansion of
observables in powers of $m_u,m_d,m_s$ around the $N_f=3$
chiral limit (meaning 3 massless flavours):
\begin{equation}
N_f=3:\qquad m_u=m_d=m_s=0\,,
\end{equation}

\emph{ii)} it is sufficiently heavy
to induce significant changes in order parameters
from the $N_f=3$ chiral limit to
the $N_f=2$ chiral limit (meaning 2 massless flavours):
\begin{equation}
N_f=2:\qquad m_u=m_d=0 \qquad m_s {\rm\ physical}\,,
\end{equation}

\emph{iii)} it is too light to suppress efficiently
loop effects of massive $\bar{s}s$ pairs (contrary to $c,b,t$ quarks).

These three arguments suggest that $\bar{s}s$ sea-pairs may play
a significant role in chiral dynamics leading to different behaviours
of QCD in $N_f=2$ and $N_f=3$ chiral limits. Then, chiral order parameters
such as the quark condensate and the pseudoscalar decay constant:
\begin{equation}
\Sigma(N_f)=-\lim_{N_f} \langle\bar{u}u\rangle\,, \qquad \qquad
F^2(N_f)=\lim_{N_f} F^2_\pi\,,
\end{equation}
would have significantly different values in the two chiral limits 
($\lim_{N_f}$ denoting the chiral limit with $N_f$ massless flavours). 

The role of $\bar{s}s$-pairs in the structure of QCD vacuum
is a typical loop effect. Therefore, it should be suppressed in 
the large-$N_c$ limit, and it can be significant
only if the Zweig rule is badly violated in the vacuum (scalar) channel
$J^{PC}=0^{++}$. On general theoretical
grounds~\cite{param}, one expects $\bar{s}s$ sea-pairs to have a 
paramagnetic effect on chiral order parameters. The latter should
decrease when the strange quark mass is sent to zero : for instance,
$\Sigma(2;m_s)\geq \Sigma(2;m_s=0)$, and similarly for $F^2$. This
corresponds to
\begin{equation} \label{eq:paramag}
\Sigma(2) \geq \Sigma(3)\,, \qquad F^2(2) \geq  F^2(3)\,.
\end{equation}
However, the size of this paramagnetic suppression is not predicted.

This effect can also be discussed in terms of the Euclidean QCD Dirac operator,
more precisely of its eigenvalue spectrum (with an
appropriate weight over the gluonic configurations)~\cite{lssr,stern}.
Chiral order parameters are related to the accumulation of the lowest
eigenvalues in the thermodynamic limit. For instance,
the quark condensate can be interpreted as the average density of
eigenvalues around 0. In this language, the Zweig-rule violating effect 
due to $\bar{s}s$ pairs corresponds to multi-point correlations 
in the density of eigenvalues around 0~\cite{param}. 
The size of the paramagnetic 
suppression eq.~(\ref{eq:paramag}) depends on the importance of such
correlations which can be interpreted as 
fluctuations~\footnote{This paramagnetic effect 
should matter only for observables dominated 
by the infrared end of the Dirac spectrum such
as the quark condensate and the pseudoscalar decay constant. 
Observables unrelated to chiral symmetry (string tension,
vector sector) would hardly be affected by
this effect and could be described accurately through large-$N_c$ 
techniques.}.

Thus, it is highly desirable to extract
the size of the chiral order parameters in $N_f=2$ and $N_f=3$ limits
from experiment.
Recent data on $\pi\pi$ scattering~\cite{E865} together with older data
and numerical solutions of the Roy 
equations~\cite{ACGL} allowed us to determine
the two-flavour order parameters expressed in suitable physical 
units~\cite{pipi}:
\begin{eqnarray} \label{eq:x2}
X(2)&=&\frac{(m_u+m_d)\Sigma(2)}{F_\pi^2M_\pi^2}=0.81\pm 0.07\,,\\
\label{eq:z2}
Z(2)&=&\frac{F^2(2)}{F_\pi^2}=0.89\pm 0.03\,.
\end{eqnarray}
A different analysis of the data in ref.~\cite{E865}, with
the additional input of dispersive estimates for the scalar radius
of the pion, led to an even larger 
value of $X(2)$~\cite{CGL}. In any case, $X(2)$ and $Z(2)$ 
are close to 1, so that corrections related to
$m_u,m_d\neq 0$ (while $m_s$ remains at its physical value) have
no significant impact on the low-energy behaviour of QCD. 
In turn, two-flavour Chiral Perturbation Theory ($\chi$PT)~\cite{chpt2},
which consists in an expansion in powers of $m_u$ and $m_d$
around the $N_f=2$ chiral limit, should suffer from no particular
problems of convergence. Indeed,
its two $O(p^2)$ low-energy constants $F^2(2)$ and $\Sigma(2)$ are
dominant in the expansions of the decay constant and mass of the pion. 

Unfortunately, two-flavour $\chi$PT~\cite{chpt2}  
deals only with dynamical pions in a very limited range of energy. 
In order to include $K$- and $\eta$-mesons dynamically
and extend the energy range of interest, one must use
three-flavour $\chi$PT~\cite{chpt3}
where the expansion in the three light quark masses starts 
around the $N_f=3$ vacuum $m_u=m_d=m_s=0$.
From the above discussion, large vacuum fluctuations of $\bar{s}s$
pairs should have a dramatic effect on $N_f=3$ chiral expansions.
The leading-order (LO) term, which depends on the $O(p^2)$ low-energy
constants $F^2(3)$ and $\Sigma(3)$, would be damped. On the other
hand, next-to-leading-order (NLO) corrections could be enhanced, in
particular those related to Zweig-rule violation in the scalar sector.
For instance, the Gell-Mann--Oakes--Renner relation would not be
saturated by its LO term and would receive sizeable numerical
contributions from terms counted as NLO in the chiral counting.

We called unstability of the expansion such a numerical 
competition between terms of different chiral counting.
A na\"{\i}ve argument based on resonance saturation suggests that
higher orders in the chiral expansion should be suppressed by powers 
of $(M_\pi/\Lambda_H)^2$. However, such an argument does not apply
to a leading-order contribution proportional to $\Sigma(3)$~: there is no 
resonance that could saturate the quark condensate. We expect
therefore to encounter three-flavour chiral 
expansions with a good overall convergence:
\begin{equation} \label{eq:conv}
A=A_{LO} + A_{NLO} + A\delta A\,, \qquad\qquad  \delta A\ll 1\,,
\end{equation}
but the numerical balance between the leading order $A_{LO}$ and 
the next-to-leading order $A_{NLO}$ depends on 
the importance of vacuum fluctuations. 

At the level of $O(p^4)$ $N_f=3$ chiral perturbation theory,
the size of the vacuum fluctuations is encoded in the low-energy constants
(LECs) $L_4$ and $L_6$ whose values remain largely unknown. For a long
time, one set them to 0 at an arbitrary hadronic scale 
(typically the $\eta$-mass)
assuming that the Zweig rule held in the scalar sector. 
More recent but indirect analyses based on dispersive 
methods~\cite{uuss,uuss2,roypika} 
suggest values of $L_4$ and $L_6$ which look quite modest but are enough
to drive the three-flavour order parameters $\Sigma(3)$ and $F^2(3)$ 
down to half of their two-flavour counterparts $\Sigma(2)$ and
$F^2(2)$. Obvioulsy, these indirect hints of sizeable vacuum fluctuations
call for a more direct confirmation.

Unstable $N_f=3$ chiral expansions ($A_{LO}\sim A_{NLO}$)
require a more careful treatment than in two-flavour $\chi$PT where
such unstabilities do not occur. For instance, it would be wrong 
to believe that the chiral expansion of $1/A$
converges nicely~\footnote{This would be equivalent to claim that
$1/(1+x)\simeq 1-x$ is a reasonable approximation for $x=O(1)$.}.
This might induce the observed problems of convergence in current two-loop 
computations~\cite{bijnens,bijnens2}~: the latter treat the fluctuations encoded
in $L_4$ and $L_6$ as small and are not designed to 
cope with a large violation of the Zweig rule in the scalar sector,
leading to unstabilities of the chiral series.

In a previous work~\cite{resum}, we proposed 
a framework to deal with chiral expansions in the case of large
fluctuations, by picking up a subset of observables with (hopefully) good
convergence properties and resumming the fluctuation
terms containing the Zweig-rule violating LECs $L_4$ and $L_6$. 
This framework includes consistently the
alternatives of large and small vacuum fluctuations.

Obviously, there is a price to pay for this extension: 
some usual $O(p^4)$ relations cannot be exploited anymore, 
because of our ignorance about their convergence.
Let us comment on a few novelties in our framework:
\begin{itemize}
\item Observables with a good convergence, eq.~(\ref{eq:conv}),
form a linear space, which we identify with connected QCD correlators
away from kinematic singularities. This choice promotes
$F_P^2$ and $F_P^2M_P^2$ ($P=\pi,K,\eta$) : LO and NLO may compete, 
but there should be only a tiny contribution form NNLO and higher. On the contrary, 
the chiral expansion of $M_P^2$ (ratio of the former quantities) 
may exhibit a bad convergence. 

In principle, two-loop computations could provide a check of these assumptions.
However, it is difficult to exploit available analyses~\cite{bijnens,bijnens2}
for two reasons: \emph{i)} in the algebraic expressions,
low-energy constants are traded for physical quantities assuming that
the chiral expansions of the latter are dominated numerically by the LO contribution,
\emph{ii)} the numerical results rely on a specific model of resonance saturation for
the $O(p^6)$ LECs~\footnote{For instance, in the resonance models used in 
refs.~\cite{bijnens,bijnens2}, $SU(3)$ breaking in quark masses is taken into account through 
a single constant in the vector sector ($f_\chi$), and it is neglected in 
the scalar sector ($d_m=0$).}. The dependence on these two assumptions should be assessed before
any definite conclusion can be drawn from two-loop computations~\footnote{Keeping in mind these issues, it remains an interesting exercise to study the convergence 
of two-loop computations for masses and decay constants. In table 2 in 
ref.~\cite{bijnens2}, four sets are considered which yield rather
different results for the overall convergence of $F_P$, $M_P^2$, $F_P^2$ and $F_P^2M_P^2$ ($P=\pi,K$). Set B exhibits small NNLO terms for $F_P$, $M_P^2$ and $F_P^2M_P^2$, but not
for $F_P^2$. In Fit 10, $F_P$, $F_P^2$, $F_P^2M_P^2$ follow eq.~(\ref{eq:conv}), 
but not $M_P^2$. For Set A and C, $F_P$ and $M_P^2$ converge, whereas $F_P^2$ and 
$F_P^2M_P^2$ suffer from large NNLO corrections.}. Obviously, if eq.~(\ref{eq:conv})
were not followed by $F_P^2$ and $F_P^2M_P^2$, but rather by other combinations of $F_P$ and $M_P$, 
some of our conclusions might be modified, but this task is beyond the scope of the present paper.

\item The three-flavour quark condensate and pseudoscalar decay constant
  expressed in physical units:
\begin{equation}
X(3)=\frac{2m\Sigma(3)}{F_\pi^2 M_\pi^2}\,, \qquad \qquad
Z(3)=\frac{F^2(3)}{F_\pi^2}\,,
\end{equation}
are free parameters. Constraints come from the vacuum
stability and the paramagnetic inequalities~(\ref{eq:paramag}):
\begin{equation}
0\leq X(3) \leq X(2)\,, \qquad \qquad 0 < Z(3) \leq Z(2)\,,
\end{equation}
where the values of $X(2)$ and $Z(2)$ have been determined 
from experiment, see eqs.~(\ref{eq:x2}-(\ref{eq:z2}).
\item The quark mass ratio $r=m_s/m$ ($m=m_u=m_d$) cannot be fixed 
from $M_K^2/M_\pi^2$ since we do not control the convergence
of its three-flavour chiral expansion. $r$ becomes a free parameter which
can vary in the range:
\begin{equation}
r_1= 2\frac{F_K M_K}{F_\pi M_\pi}-1 \sim 8 \leq r \leq 
r_2= 2\frac{F_K^2 M_K^2}{F_\pi^2 M_\pi^2}-1 \sim 36\,.
\end{equation}
The chiral expansions of $F_\pi^2M_\pi^2$ and
$F_K^2 M_K^2$ lead to a correlation between $r$ and $X(2)$~\cite{param}.
Experimentally, the
 analysis of $\pi\pi$ scattering phase shifts~\cite{E865} shows
that $r>14$ (95\% CL), slightly favouring values between 20 and 
25~\cite{resum}.
\item The agreement of the pseudoscalar spectrum with the 
Gell-Mann--Okubo formula requires a fine tuning of $L_7$. Let us remark
though that this fine tuning exists even in the case of a dominant 
three-flavour quark condensate and small vacuum fluctuations~\cite{resum}.
\item One cannot determine LECs or combinations of LECs through ratios
of observables. For instance, one should not use $F_K/F_\pi$ to
determine $L_5$ because we do not know if the chiral expansion of 
$F_K/F_\pi$ converges at all.
\end{itemize}

There are some prospects of probing experimentally the three-flavour
sector and in particular the size of vacuum fluctuations 
through $\pi K$ scattering~\cite{roypika}. 
Unfortunately, the current data are
not precise enough to draw any definite conclusion. On the other hand,
recent progress achieved by lattice unquenched
simulations~\cite{lattrev} 
makes them an interesting field to investigate 
the size of vacuum fluctuations. For 
instance, lattice practitionners can vary very easily the value of the
quark masses.

A comment is in order at this point.
From large-$N_c$ considerations, it is often assumed that 
quark-loop effects are not significant, so that simulations
with only two dynamical flavours, or even none (quenched case), could
be good approximations to real QCD. We are precisely questionning
this assumption in the case of observables related to chiral symmetry
breaking. Thus, it is mandatory to perform unquenched simulations 
with three dynamical flavours in order to probe 
strange sea-quark effects on the pattern of chiral symmetry breaking,

One could think of investigating
directly the correlations of Dirac eigenvalues, or the $m_s$-dependence
of the quark condensate and pseudoscalar decay constant. However, this
might prove rather challenging tasks since they require simulations with
three dynamical very light quarks.
In this paper, we investigate another way of probing the size
of vacuum fluctuations through the spectrum of the theory.
Indeed, this less immediate approach is easier to follow with
current lattice simulations. We believe that this exercise might also
be useful to illustrate some subtleties arising in 
three-flavour chiral extrapolations when vacuum fluctuations of
$s\bar{s}$ pairs are not negligible.

We are not going to address the issue of discretisation, which depends on the
specific implementation of the lattice action. We focus on chiral 
extrapolations with potentially large vacuum fluctuations and on the impact of
finite-volume corrections. For definiteness,
we work in the isospin limit $m_u=m_d=m$ and
we consider a lattice simulation with (2+1) flavours : 
two flavours are set to a common mass $\tilde{m}$, whereas the third one 
is kept at the same mass as the physical strange quark $m_s$.
Each quantity $X$ observed in the physical situation $(m,m,m_s)$ has
a lattice counterpart $\tilde{X}$ for $(\tilde{m},\tilde{m},m_s)$.

We introduce the notation:
\begin{eqnarray}
q=\frac{\tilde{m}}{m_s}\,, &\quad& r=\frac{m_s}{m}\,,\\
X(3)=\frac{2m\Sigma(3)}{F_\pi^2M_\pi^2}\,,
 & \quad & Z(3)=\frac{F^2(3)}{F_\pi^2}\,,\\
Y(3)&=&\frac{X(3)}{Z(3)}=\frac{2mB_0}{M_\pi^2}\,,
\end{eqnarray}
where $\Sigma(3)$ and $F(3)$ are (the absolute value of) the quark
condensate and the pion decay constant in the $N_f=3$ chiral limit,
and $B_0=\Sigma(3)/F(3)^2$. We take the following values for
the masses and decay constants: $F_\pi=92.4$ MeV, $F_K/F_\pi=1.22$,
$M_\pi=139.6$ MeV, $M_K=493.7$ MeV, $M_\eta=547$ MeV.

\section{Vacuum fluctuations at infinite volume}

\subsection{Bare expansions of masses and decay constants}

\label{sec:bareinf}

We start by considering the impact of vacuum 
fluctuations on the spectrum in the limit $L\to\infty$.
Since large vacuum fluctuations are allowed, the problems
highlighted in the introduction might arise. Therefore,
we have to define the appropriate observables to consider and
the treatment of their chiral expansion. 
We follow the procedure advocated in~\cite{resum}, reexpressing
it in an equivalent way convenient for our purposes:
\begin{enumerate}
\item Consider a subset of observables that are assumed to have a 
good overall convergence -- we call them ``good observables''. They
must form a linear space, which we choose to be that of 
connected QCD correlators
(of vector/axial currents and their divergences) as functions
of external momenta, away from any kinematic singularities. 
This rule selects in particular $F_P^2$ and $F_P^2M_P^2$.
\item Take each observable and write its NLO chiral expansion.
\item In theses formulae, reexpand the physical quantities
(masses, decay constants\ldots)
in powers of quark masses wherever the resulting
dependence on the latter is polynomial. Keep the physical
quantities only in the nonanalytic terms 
(unitarity cuts, logarithmic divergences\ldots).
The result is called ``bare expansion''.
\item In the bare expansions, reexpress $O(p^2)$ 
and $O(p^4)$ LECs in terms of $X(3)$, $Z(3)$ and $r$
using exact Ward identities for pseudoscalar masses and decay constants.
\end{enumerate}

The first step is straightforward : one has to compute 
the NLO chiral expansion of the masses and decay constants of 
Goldstone bosons in an infinite volume with 
the quark masses $m_u=m_d=\tilde{m}$ and $m_s$. The expressions
can be obtained directly from ref.~\cite{chpt3}.
Then, we must reexpand the physical quantities
occuring in these formulae in terms of the quark masses wherever
the dependence is analytic. In the case of $\tilde{F}_P^2$ and 
$\tilde{F}_P^2 \tilde{M}_P^2$, the only issue lies in
tadpole contributions such as:
\begin{equation}
\frac{\tilde{M}_\pi^2}{16\pi^2}\log\frac{\tilde{M}_\pi^2}{\mu^2}\,.
\end{equation}
To obtain the bare expansions associated with $\tilde{F}_P^2$ and 
$\tilde{F}_P^2 \tilde{M}_P^2$, we keep the physical masses only
in the (nonanalytic) logarithm but we expand the front factor
in powers of quark masses. Hence, at the chiral order we are working, 
the contribution from the pion tadpole 
has the following bare expansion:
\begin{equation}\label{eq:tadredef}
\frac{2{\tilde m}B_0}{16\pi^2}\log\frac{{\tilde M}_\pi^2}{\mu^2}\,.
\end{equation}
In the bare expansion, the polynomial structure in 
the quark masses is explictly
displayed, so that we can keep track of the relative contribution
of the LO, NLO, NNLO$\ldots$ terms. 

The resulting bare expansions for the decay constants (third
step of our procedure) can be expressed in the following way:
\begin{eqnarray} \label{eq:fpilatt}
&&\frac{\tilde{F}_\pi^2}{F_\pi^2} =
  Z(3)+\frac{M_\pi^2}{F_\pi^2}qrY(3)
           \left[8\left(\frac{1}{q}+2\right)L_4^r+8L_5^r\right]\\
\nonumber
&&\qquad      -\frac{M_\pi^2}{F_\pi^2}\frac{1}{32\pi^2}qrY(3)
           \left[4\log\frac{\tilde{M}_\pi^2}{\mu^2}
            +\left(\frac{1}{q}+1\right)\log\frac{\tilde{M}_K^2}{\mu^2}\right]\\
&&\qquad
 +\frac{\tilde{F}_\pi^2}{F_\pi^2}\tilde{e}_\pi\,,
\nonumber
\\
&&\frac{\tilde{F}_K^2}{F_\pi^2} =
  Z(3)\\
\nonumber
&&\qquad
+\frac{M_\pi^2}{F_\pi^2}qrY(3)
           \left[8\left(\frac{1}{q}+2\right)L_4^r
                     +4\left(\frac{1}{q}+1\right)L_5^r\right]\\
\nonumber
&&\qquad      -\frac{M_\pi^2}{F_\pi^2}\frac{1}{32\pi^2}qrY(3)
      \Bigg[\frac{3}{2}\log\frac{\tilde{M}_\pi^2}{\mu^2}
\nonumber
\\
&&\qquad\qquad
            +\frac{3}{2}
         \left(\frac{1}{q}+1\right)\log\frac{\tilde{M}_K^2}{\mu^2}
            +\frac{1}{2}\left(\frac{2}{q}+1\right)
                 \log\frac{\tilde{M}_\eta^2}{\mu^2}
           \Bigg]\nonumber
\\
&&\qquad
 +\frac{\tilde{F}_K^2}{F_\pi^2}\tilde{e}_K\,,
\nonumber
\\
&&\frac{\tilde{F}_\eta^2}{F_\pi^2} =
  Z(3)]\\
\nonumber
&&\qquad
+\frac{M_\pi^2}{F_\pi^2}qrY(3)
           \left[8\left(\frac{1}{q}+2\right)L_4^r
                     +\frac{8}{3}\left(\frac{2}{q}+1\right)L_5^r\right]\\
&&\qquad       -\frac{M_\pi^2}{F_\pi^2}\frac{1}{32\pi^2}qrY(3)
           \times
  3\left(\frac{1}{q}+1\right)\log\frac{\tilde{M}_K^2}{\mu^2}
+\frac{\tilde{F}_\eta^2}{F_\pi^2}\tilde{e}_\eta\,,
\nonumber
\end{eqnarray}
where $\tilde{e}_P$ are NNLO remainders of $O(m_q^2)$ ($m_q$ denotes
either $m_s$ or $\tilde{m}$). We have divided by the physical value
of $F_\pi^2$ in order to deal with dimensionless quantities.

In a similar way, we obtain the bare expansions of the masses:
\begin{eqnarray}
&&\frac{\tilde{F}_\pi^2\tilde{M}_\pi^2}{F_\pi^2M_\pi^2} =
  qr\Bigg\{X(3)\\
&&\qquad
+\frac{M_\pi^2}{F_\pi^2}qr[Y(3)]^2
           \left[16\left(\frac{1}{q}+2\right)L_6^r+16L_8^r\right]
\nonumber\\
&&\qquad      -\frac{M_\pi^2}{F_\pi^2}\frac{1}{32\pi^2}qr[Y(3)]^2
           \Bigg[3\log\frac{\tilde{M}_\pi^2}{\mu^2}
\nonumber\\
&&\qquad\qquad\qquad
            +\left(\frac{1}{q}+1\right)\log\frac{\tilde{M}_K^2}{\mu^2}
            +\frac{1}{9}\left(\frac{2}{q}+1\right)
                   \log\frac{\tilde{M}_\eta^2}{\mu^2}
           \Bigg]\Bigg\}\nonumber\\
&&\qquad
 + \frac{\tilde{F}_\pi^2\tilde{M}_\pi^2}{F_\pi^2M_\pi^2}\tilde{d}_\pi\,,
\nonumber
\\
&&\frac{\tilde{F}_K^2\tilde{M}_K^2}{F_\pi^2M_\pi^2} =
  \frac{qr}{2}\left(\frac{1}{q}+1\right)
    \Bigg\{X(3)
\\
&&\qquad
+\frac{M_\pi^2}{F_\pi^2}qr[Y(3)]^2
           \left[16\left(\frac{1}{q}+2\right)L_6^r
                     +8\left(\frac{1}{q}+1\right)L_8^r\right]
\nonumber\\
&&\qquad
  -\frac{M_\pi^2}{F_\pi^2}\frac{1}{32\pi^2}qr[Y(3)]^2
           \Bigg[\frac{3}{2}\log\frac{\tilde{M}_\pi^2}{\mu^2}\nonumber\\
&&\qquad\qquad
            +\frac{3}{2}
                \left(\frac{1}{q}+1\right)\log\frac{\tilde{M}_K^2}{\mu^2}
            +\frac{5}{18}\left(\frac{2}{q}+1\right)
                   \log\frac{\tilde{M}_\eta^2}{\mu^2}
           \Bigg]\Bigg\}\nonumber\\
&&\qquad
 +\frac{\tilde{F}_K^2\tilde{M}_K^2}{F_\pi^2M_\pi^2}\tilde{d}_K\,,
\nonumber
\\ \label{eq:fetametalatt}
&&\frac{\tilde{F}_\eta^2\tilde{M}_\eta^2}{F_\pi^2M_\pi^2} =
  qr \Bigg\{
     \frac{1}{3}\left(\frac{2}{q}+1\right)\\
&&\qquad\times       \Bigg(X(3)+\frac{M_\pi^2}{F_\pi^2}qr[Y(3)]^2\nonumber\\
&&\qquad\qquad\qquad\times 
           \left[16\left(\frac{1}{q}+2\right)L_6^r
                     +\frac{16}{3}\left(\frac{2}{q}+1\right)L_8^r\right]
\nonumber\\
&&\qquad\qquad      -\frac{M_\pi^2}{F_\pi^2}\frac{1}{32\pi^2}qr[Y(3)]^2
\nonumber\\
&&\qquad\qquad\times 
           \left[
              2\left(\frac{1}{q}+1\right)\log\frac{\tilde{M}_K^2}{\mu^2}
            +\frac{4}{9}\left(\frac{2}{q}+1\right)
                   \log\frac{\tilde{M}_\eta^2}{\mu^2}
           \right]\Bigg)
\nonumber
\\
&&\quad -\frac{M_\pi^2}{F_\pi^2}\frac{1}{32\pi^2}qr[Y(3)]^2
  \Bigg[
              \log\frac{\tilde{M}_\pi^2}{\mu^2}\nonumber
\\
 &&\qquad\qquad
            -\frac{1}{3}
                  \left(\frac{1}{q}+1\right)\log\frac{\tilde{M}_K^2}{\mu^2}
           -\frac{1}{9}\left(\frac{2}{q}+1\right)
                   \log\frac{\tilde{M}_\eta^2}{\mu^2}
           \Bigg]
\nonumber\\
&&\quad +\frac{32}{9}\left(\frac{1}{q}-1\right)^2
                  \frac{M_\pi^2}{F_\pi^2}qr[Y(3)]^2[3L_7+L_8^r]
\Bigg\}
 +\frac{\tilde{F}_\eta^2\tilde{M}_\eta^2}{F_\pi^2M_\pi^2}\tilde{d}_\eta\,,
\nonumber
\end{eqnarray}
where $\tilde{d}_P$ are NNLO remainders of $O(m_q^2)$.
 We have divided by the physical value
of $F_\pi^2 M_\pi^2$ in order to deal with dimensionless quantities.

\subsection{Expression of $O(p^4)$ LECs}

We must know perform the fourth step of our procedure.
As shown in refs.~\cite{param,uuss,uuss2,resum}, one can reexpress
the $O(p^4)$ LECs $L_4,L_5,L_6,L_8$ in terms
of $[r,X(3),Z(3)]$ (and NNLO remainders) using the four chiral 
expansions of $F_P^2$ and $F_P^2M_P^2$ ($P=\pi,K$) in the physical case. 
These quantities, related to two-point functions of 
axial/vector currents and of their divergences at vanishing
momentum transfer, are expected to 
have small NNLO remainders. One obtains~\cite{resum}:
\begin{eqnarray} 
\label{eq:delta4}
Y(3)\Delta L_4 &=& \frac{1}{8(r+2)}\frac{F_\pi^2}{M_\pi^2}
  [1-\eta(r)-Z(3)-e]\,,
\\ 
\label{eq:delta5}
Y(3)\Delta L_5 &=& \frac{1}{8}\frac{F_\pi^2}{M_\pi^2}
  [\eta(r)+e']\,, 
\\
\label{eq:delta6}
Y^2(3)\Delta L_6 &=& \frac{1}{16(r+2)}\frac{F_\pi^2}{M_\pi^2}
  [1-\epsilon(r)-X(3)-d]\,,
\\ 
\label{eq:delta8}
Y^2(3)\Delta L_8 &=& \frac{1}{16}\frac{F_\pi^2}{M_\pi^2}
  [\epsilon(r)+d']\,.
\end{eqnarray}
$\Delta L_i=L_i^r(\mu)-\hat{L}_i(\mu)$ combine the 
(renormalized and quark-mass independent) 
constants $L_{4,5,6,8}$ and chiral logarithms so that they are
independent of the renormalisation scale $\mu$: 
\begin{eqnarray}
\label{eq:l4}
32\pi^2\hat{L}_4(\mu)
          &=& \frac{1}{8}
          \log\frac{M_K^2}{\mu^2}
  -\frac{1}{8(r-1)(r+2)}\\
&&\times
  \left[(4r+1)\log \frac{M_K^2}{M_\pi^2} 
      + (2r+1)\log \frac{M_\eta^2}{M_K^2}  \right]       
\nonumber\,,\\
\label{eq:l5}
32\pi^2 \hat{L}_5(\mu) &=& \frac{1}{8}
    \left[\log\frac{M_K^2}{\mu^2}+2\log\frac{M_\eta^2}{\mu^2}\right]
    \\
&&\quad   +\frac{1}{8(r-1)}
    \left[3\log\frac{M_\eta^2}{M_K^2}+5\log\frac{M_K^2}{M_\pi^2}\right]
\nonumber\,,\\
\label{eq:l6}
32\pi^2\hat{L}_6(\mu) &=& \frac{1}{16}\left[
       \log\frac{M_K^2}{\mu^2} 
       + \frac{2}{9}\log \frac{M_\eta^2}{\mu^2}
         \right]  \\
 &&  -\frac{1}{16}\frac{r}{(r+2)(r-1)} \left[ 3 \log
          \frac{M_K^2}{M_\pi^2}  + \log \frac{M_\eta^2}{M_K^2} \right]
\nonumber\,, \\
\label{eq:l8}
32\pi^2 \hat{L}_8(\mu) &=& \frac{1}{16}
    \left[\log\frac{M_K^2}{\mu^2}+\frac{2}{3}\log\frac{M_\eta^2}{\mu^2}\right]
 \\
&& +\frac{1}{16(r-1)} \left[ 3 \log
          \frac{M_K^2}{M_\pi^2}  + \log \frac{M_\eta^2}{M_K^2} \right]
\nonumber\,.
\end{eqnarray}

The right hand-side of eqs.~(\ref{eq:delta4})-(\ref{eq:delta8})
involves the $r$-dependent functions:
\begin{eqnarray} \label{eq:funcr1}
\epsilon(r) &=& 2\frac{r_2-r}{r^2-1}, \quad
r_2 = 2\left(\frac{F_KM_K}{F_{\pi}M_{\pi}}\right)^2 -1\sim 36\,,
\\
\label{eq:funcr2}
\eta(r)&=&\frac{2}{r-1}\left(\frac{F_K^2}{F_\pi^2}-1\right)\,,
\end{eqnarray}
whereas $d,d'$ and $e,e'$ are combinations of
NNLO remainders associated with the chiral expansions of $\pi,K$ masses
and decay constants respectively. These remainders
should be small, and we are going to neglect them for all numerical 
estimates in the following. 

We stress that eqs.~(\ref{eq:delta4})-(\ref{eq:delta8}) are nothing
more than a convenient reexpression of the $N_f=3$ chiral expansions
of $F_\pi^2, F_K^2, F_\pi^2 M_\pi^2, F_K^2M_K^2$. The latter
can be easily recovered through linear combinations of 
eqs.~(\ref{eq:delta4})-(\ref{eq:delta8}). For instance, the chiral
expansion of $F_\pi^2$ [$F_\pi^2M_\pi^2$] is obtained when one
combines eqs.~(\ref{eq:delta4})-(\ref{eq:delta5}) 
[eqs.~(\ref{eq:delta6})-(\ref{eq:delta8})] to eliminate $\eta(r)$ 
[$\epsilon(r)$].

\subsection{Numerical results}

To exploit the previous formulae, we have to fix the values of
$r,X(3),Z(3)$. Since these parameters are only
weakly constrained by experimental data, we will vary them
in the ranges $r=20,30$, $Z(3)=0.4,0.8$ and $X(3)=0.2,0.4,0.8$. In each
``physical'' situation, we study how the masses and decay constants of
the simulated $\pi$ and $K$ vary with $q$, neglecting all NNLO
remainders for the moment. 

A slight computational difficulty arises: in 
eqs.~(\ref{eq:fpilatt})-(\ref{eq:fetametalatt}),
the logarithmic NLO corrections involve the values of the
simulated masses $\tilde{M}_P^2$. We could solve iteratively the system
of equations to determine these masses. However, two iterations turn out
to achieve a sufficient numerical accuracy. This corresponds to an
easier procedure: \emph{i)} compute $\hat{M}_P^2$ defined from
eqs.~(\ref{eq:fpilatt})-(\ref{eq:fetametalatt}) with
$\tilde{M}_P^2$ replaced by $M_P^2$ on their right-hand side, \emph{ii)}
consider eqs.~(\ref{eq:fpilatt})-(\ref{eq:fetametalatt}) again, with
$\tilde{M}_P^2$ replaced by $\hat{M}_P^2$ on their right-hand side, \emph{iii)}
the resulting values for the pseudoscalar masses are 
equal to $\tilde{M}_P^2$ up to a tiny error.

\begin{figure*}
\begin{center}
\hbox{
\includegraphics[width=8cm]{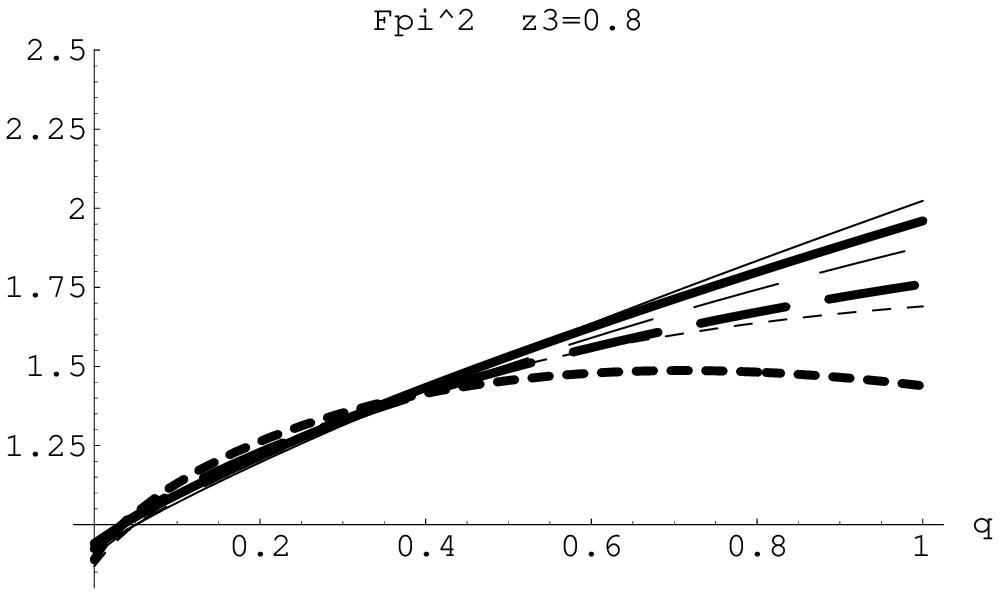}
\includegraphics[width=8cm]{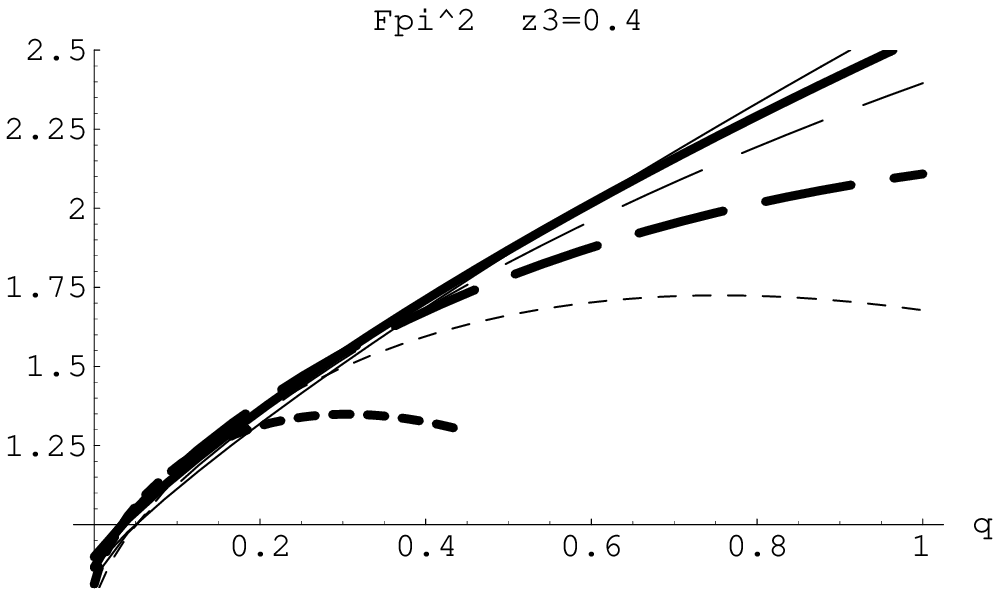}}
\hbox{
\includegraphics[width=8cm]{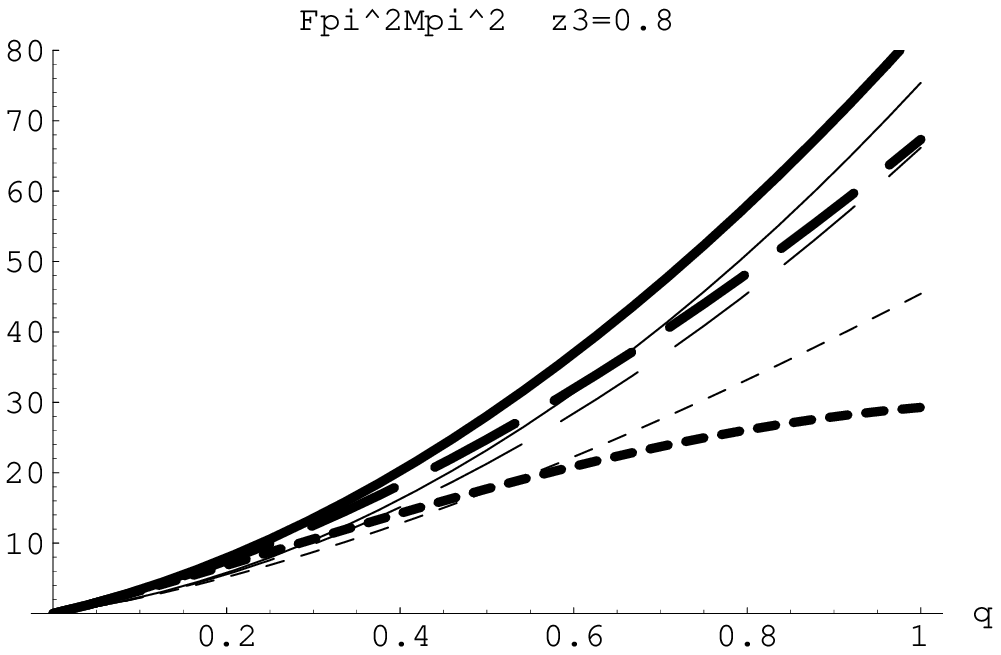}
\includegraphics[width=8cm]{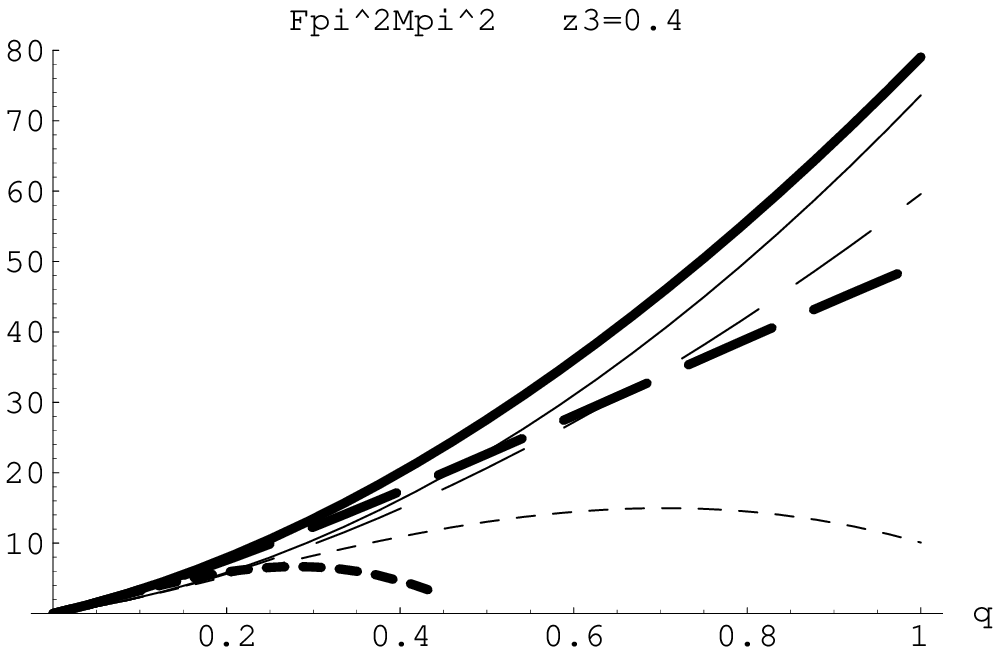}}
\hbox{
\includegraphics[width=8cm]{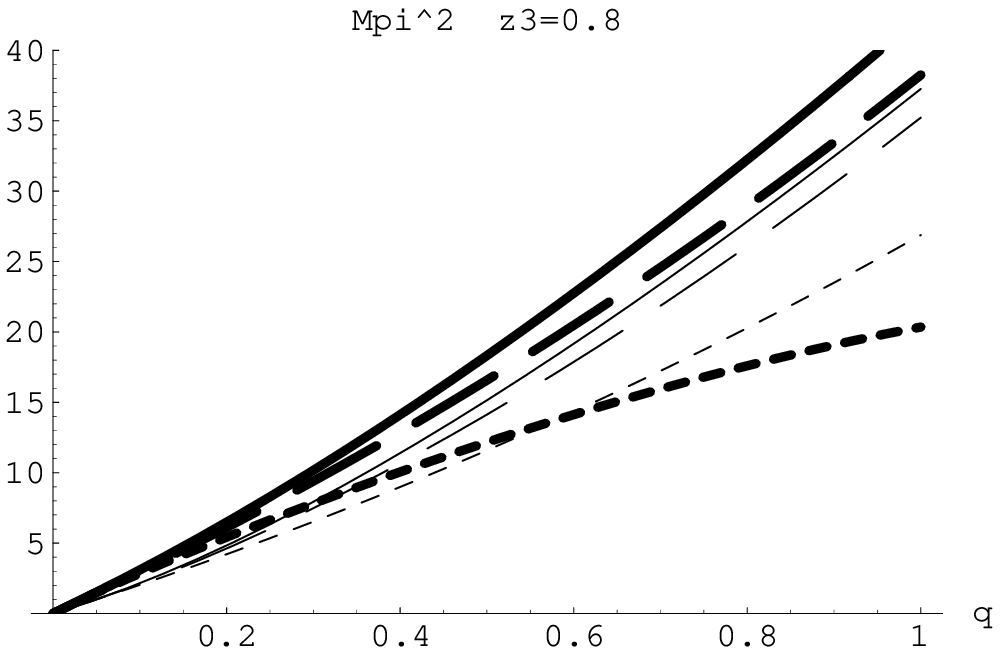}
\includegraphics[width=8cm]{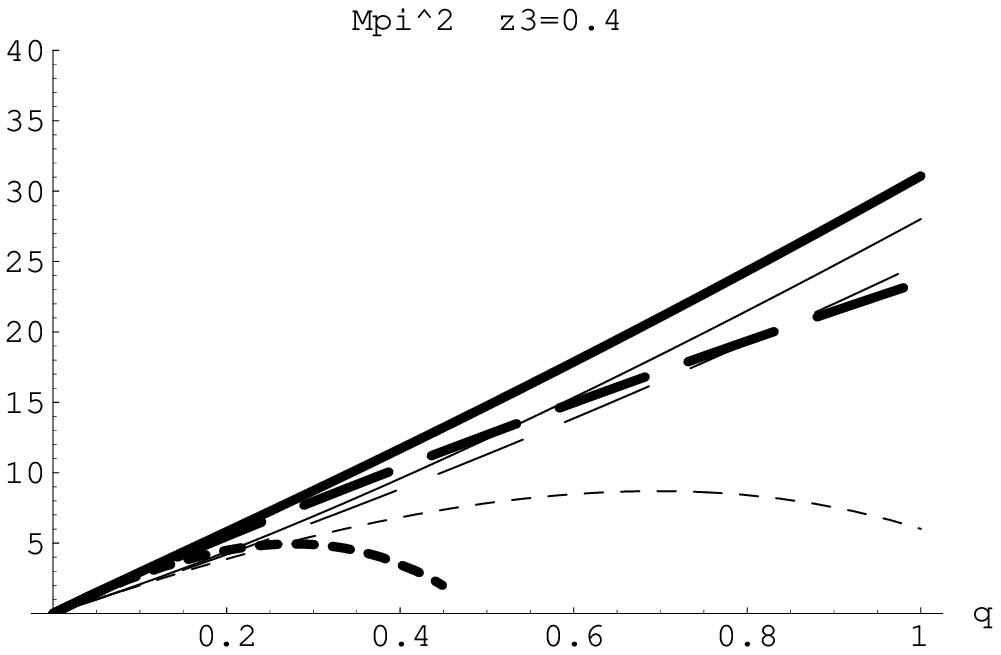}}
\end{center}
\caption{$\tilde{F}_\pi^2$, 
$\tilde{F}_\pi^2\tilde{M}_\pi^2$ and $\tilde{M}_\pi^2$ (respectively upper, 
middle and lower plots) normalized to their physical values as functions of
$q=\tilde{m}/m_s$. 
The first (second) column corresponds to $Z(3)=0.8$ (0.4).
On each plot, full, long-dashed and dashed curves correspond respectively
to $X(3)=0.8, 0.4, 0.2$. Thick (thin) lines are drawn for $r=30$ (20).
All NNLO remainders are neglected.}
\label{fig:pinofinvol}
\end{figure*}

\begin{figure*}
\begin{center}
\hbox{
\includegraphics[width=8cm]{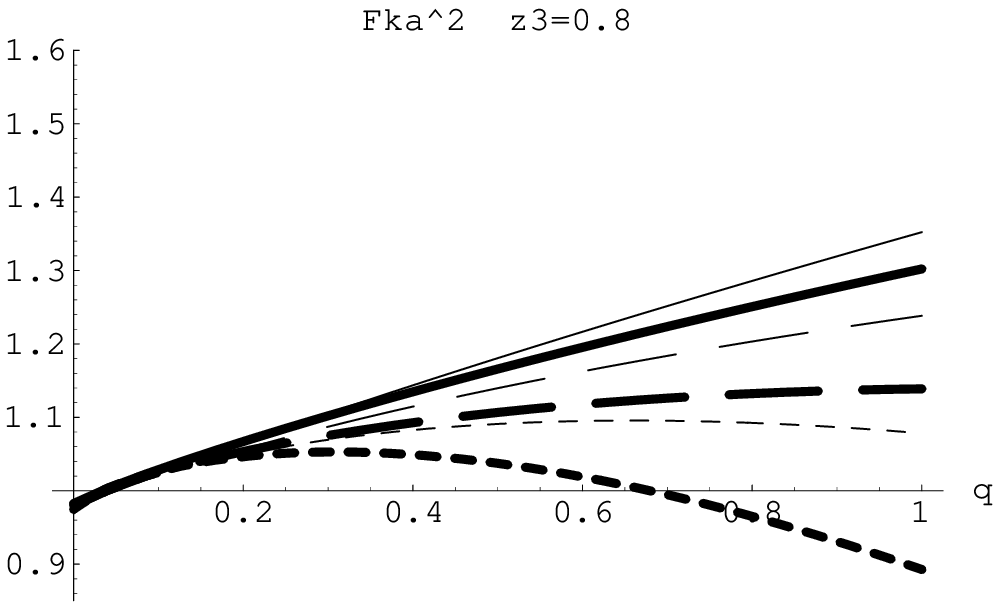}
\includegraphics[width=8cm]{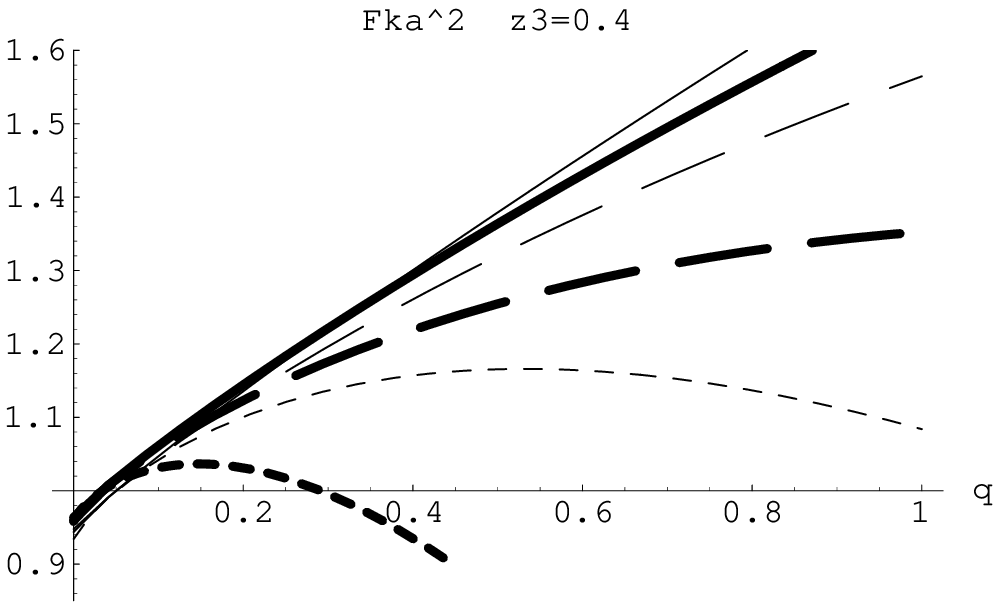}}
\hbox{
\includegraphics[width=8cm]{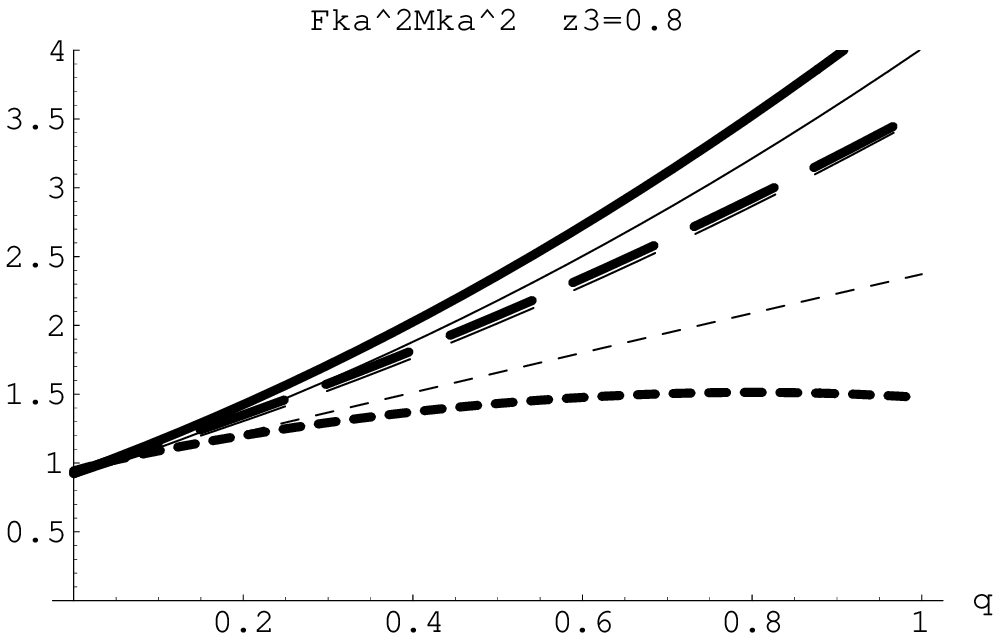}
\includegraphics[width=8cm]{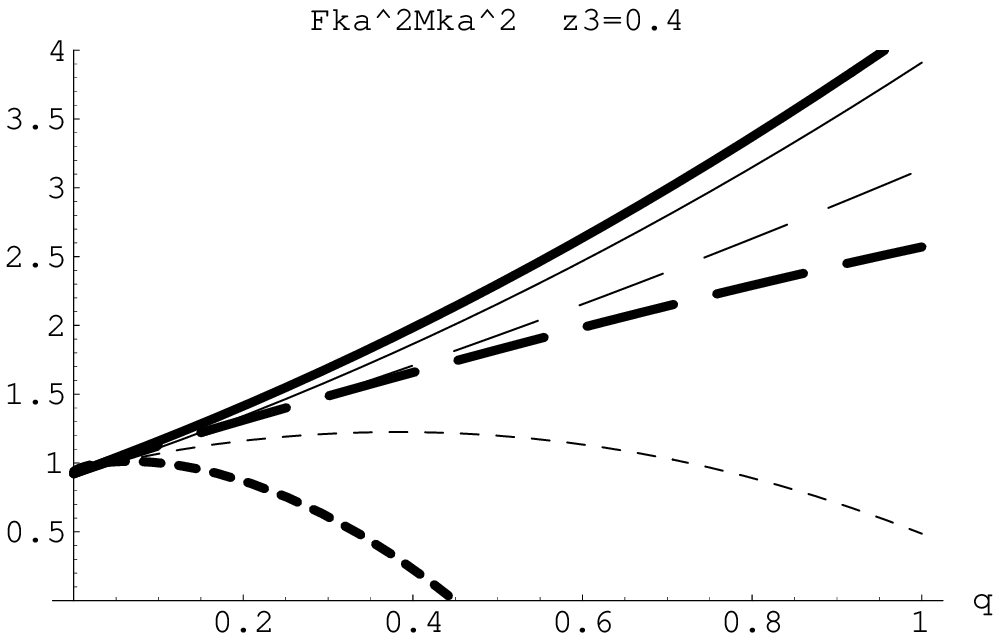}}
\hbox{
\includegraphics[width=8cm]{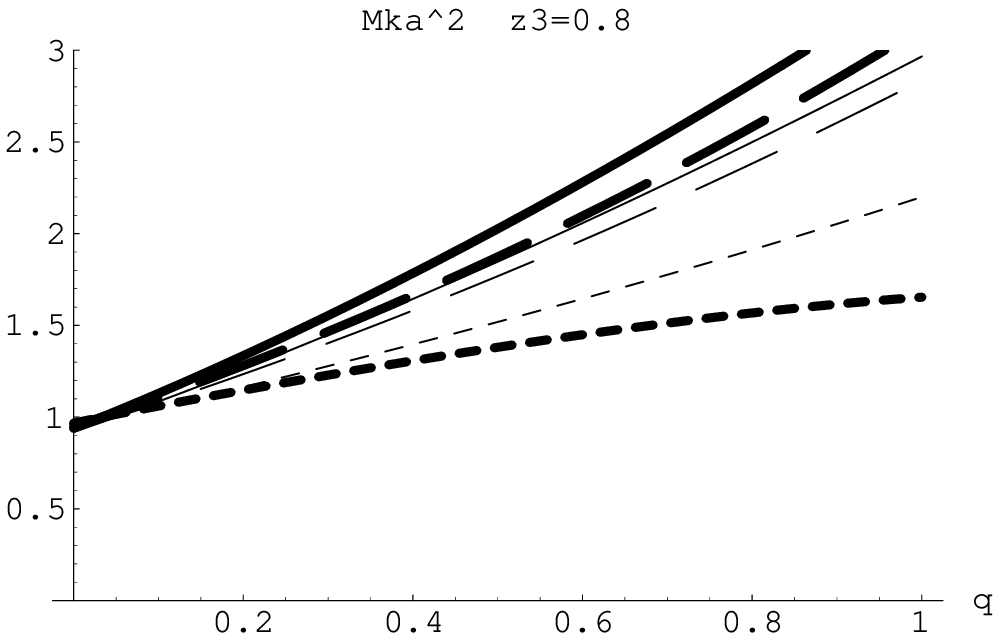}
\includegraphics[width=8cm]{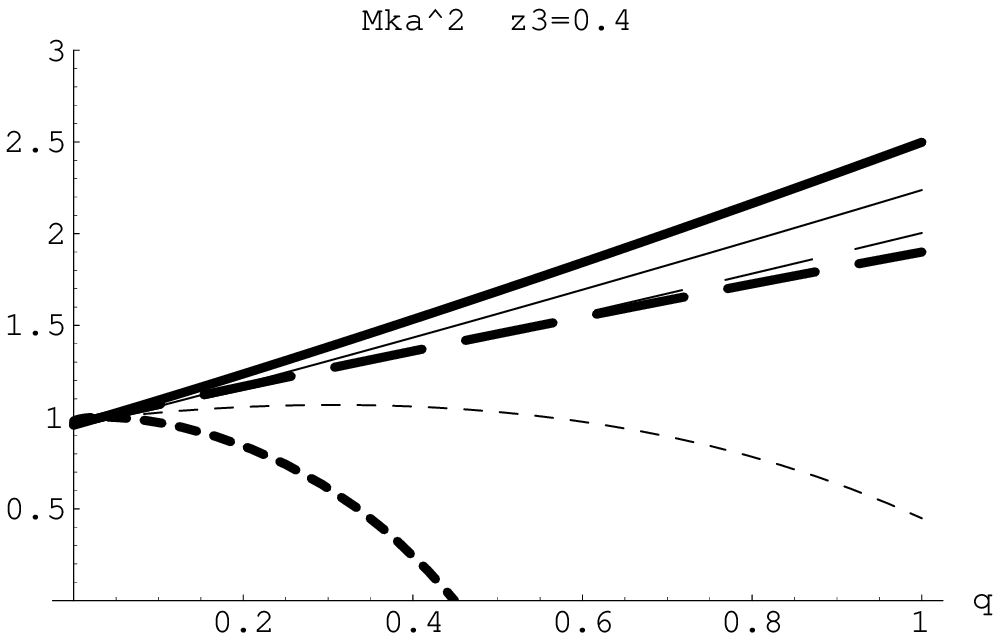}}
\end{center}
\caption{$\tilde{F}_K^2$, 
$\tilde{F}_K^2\tilde{M}_K^2$ and 
$\tilde{M}_K^2$ (respectively upper,
middle and lower plots) normalised to their physical values as functions of
$q=\tilde{m}/m_s$. 
The first (second) column corresponds to $Z(3)=0.8$ (0.4).
On each plot, full, long-dashed and dashed curves correspond respectively
to $X(3)=0.8, 0.4, 0.2$. Thick (thin) lines are drawn for $r=30$ (20).
All NNLO remainders are neglected.}
\label{fig:kanofinvol}
\end{figure*}

In figs.~\ref{fig:pinofinvol} and \ref{fig:kanofinvol}
we have plotted $\tilde{F}_P^2$, $\tilde{F}_P^2\tilde{M}_P^2$ 
and $\tilde{M}_P^2$ ($P=\pi,K$) 
as functions of $q=\tilde{m}/m_s$ in an infinite volume,
neglecting all NNLO remainders. 
For each observable, the first column corresponds to
$Z(3)=0.8$, the second one to $Z(3)=0.4$. 
On each plot, the curves 
correspond to different values of $X(3)$ [full: 0.8, long-dashed: 0.4, 
dashed: 0.2] and of $r$ [thick: $r=30$, thin: $r=20$].

A few comments are in order:
\begin{itemize}
\item Our choice of normalisation imposes that all the curves intersect at
the physical point $\tilde{m}=m$ where:
\begin{equation}
q=\frac{1}{r}\,,\qquad\qquad \frac{\tilde{X}}{X}=1\,.
\end{equation}

\item
If $Y(3)>1$, vacuum instability may occur : the pion mass becomes
negative for ``large'' masses [$q=O(1)$]. This situation would
occur if vacuum fluctuations had a more significant impact
on the decay constant than on the condensate : the first would decrease
more quickly than the latter from the physical case to the three-flavour
chiral limit. 

Actually, this situation seems unlikely. A nonvanishing $F^2$ is
equivalent to the spontaneous breakdown of chiral symmetry. Other
chiral order parameters (like $\Sigma$) may or may not vanish depending
on the breaking pattern. We expect therefore $F^2$ to be the last 
chiral order parameter to vanish, after or together with
all the other parameters (e.g., $\Sigma$), and thus $Y(3)\leq 1$. 
The analysis of some properties of the small Dirac eigenvalues 
suggests the same conclusion~\cite{lssr,stern}.

Even though this theoretical expectation has not been checked experimentally
yet, we dismiss the case $Y(3)\geq 1$ (unphysical in our opinion)
in the remainder of this article.

\item
On the plots, very small values of $X(3)$ do not necessarily correspond 
to almost vanishing pion masses. Indeed, we take into acount at least
two different sources of chiral symmetry breaking : 
quark condensation $\Sigma(3)$
and vacuum fluctuations $L_4$, $L_6$. For $X(3)=0$,
chiral symmetry breaking may still be triggered by the fluctuations encoded 
in $L_6$, and the simulated pion mass may remain rather massive 
for $\tilde{m}\neq 0$ .

\item 
The pseudoscalar masses $M_P^2$ are not the best candidates to observe 
a curvature due to chiral logarithms. The $\tilde{m}$-dependence is hardly
different from a polynomial one. In addition, $X(3)$ must be much
smaller than 1 to observe a variation in the curvature.
\end{itemize}

\section{Finite-volume effects}

\subsection{NLO chiral expansions}

The lattice simulations are performed in a finite spatial box, whereas time
is sent to infinity to single out the ground state. For sufficiently
large boxes ($L\gg 1/\Lambda_H$), the Goldstone modes remain the only
relevant degrees of freedom, whose interactions
can be described through a low-energy effective theory~\cite{GLfinvol}.
In the case of periodic boundary conditions, this effective theory 
is identical to $\chi$PT, with the same values of the 
LECs as in an infinite volume. In addition, if 
the box size is large enough compared
to the inverse Compton length of the
pion~\cite{GLfinvol,luscher,finvol},
the so-called $p$-expansion is valid and
the only difference from the infinite-volume case shows up in the
propagators of the Goldstone modes. This
affects only the tadpole logarithms in the formulae of the previous section:
\begin{equation}
\frac{\tilde{M}_P^2}{16\pi^2}
  \log\frac{\tilde{M}_P^2}{\mu^2}
 \qquad \to \qquad \frac{1}{2L^3}\sum_{\vec\ell}\frac{1}{\omega_P}\,,
\end{equation}
where:
\begin{equation}
\vec\ell\in \frac{2\pi}{L}\times Z^3\,, \qquad\qquad
\omega_P=\sqrt{\vec{\ell}^2+\tilde{M}_P^2}\,.
\end{equation}
We obtain for the decay constants:
\begin{eqnarray}
&&\frac{\tilde{F}_\pi^2}{F_\pi^2} =
  Z(3)+\frac{M_\pi^2}{F_\pi^2}qrY(3)
           \left[8\left(\frac{1}{q}+2\right)L_4^r+8L_5^r\right]\\
&&\qquad   
    -\frac{1}{4F_\pi^2L^3}
         \left[4\tilde\sigma_\pi
            +2\tilde\sigma_K\right]\,,
\nonumber
\\
&&\frac{\tilde{F}_K^2}{F_\pi^2} =
  Z(3)\\
&&\qquad +\frac{M_\pi^2}{F_\pi^2}qrY(3)
           \left[8\left(\frac{1}{q}+2\right)L_4^r
                     +4\left(\frac{1}{q}+1\right)L_5^r\right]
\nonumber\\
&&\qquad      -\frac{1}{4F_\pi^2L^3}
           \left[\frac{3}{2}\tilde\sigma_\pi
                +3\tilde\sigma_K
                +\frac{3}{2}\tilde\sigma_\eta
           \right]\,,
\nonumber
\end{eqnarray}
where:
\begin{equation} \label{eq:sigmaP}
\tilde\sigma_P=
  \sum_{\vec\ell}\frac{1}{\sqrt{\vec\ell^2+\tilde{M}_P^2}}\,.
\end{equation}

In a similar way, we obtain for the masses:
\begin{eqnarray}
&&\frac{\tilde{F}_\pi^2\tilde{M}_\pi^2}{F_\pi^2M_\pi^2} =
  qr\Bigg\{X(3) \\
&&\qquad
+\frac{M_\pi^2}{F_\pi^2}qr[Y(3)]^2
           \left[16\left(\frac{1}{q}+2\right)L_6^r+16L_8^r\right]\nonumber\\
&&\qquad      -\frac{Y(3)}{4F_\pi^2L^3}
           \left[3\tilde\sigma_\pi
            +2\tilde\sigma_K
            +\frac{1}{3}\tilde\sigma_\eta
           \right]\Bigg\}\,,
\nonumber
\\
&&\frac{\tilde{F}_K^2\tilde{M}_K^2}{F_\pi^2M_\pi^2} =
  \frac{qr}{2}\left(\frac{1}{q}+1\right)\Bigg\{X(3)\\
&&\qquad +\frac{M_\pi^2}{F_\pi^2}qr[Y(3)]^2
           \left[16\left(\frac{1}{q}+2\right)L_6^r
                     +8\left(\frac{1}{q}+1\right)L_8^r\right]
\nonumber
\\
&&\qquad      -\frac{Y(3)}{4F_\pi^2L^3}
           \left[\frac{3}{2}\tilde\sigma_\pi
            +3\tilde\sigma_K
            +\frac{5}{6}\tilde\sigma_\eta
           \right]\Bigg\}\,.
\nonumber
\end{eqnarray}

\subsection{Bare expansions at finite volume}

The above NLO chiral expansions are not bare expansions yet.
We must reexpand the physical masses
in powers of quark masses wherever the dependence on the latter is
polynomial. Therefore, we have to identify the nonanalytic pieces in
the tadpole term $\tilde\sigma$, which are a logarithm due to
ultraviolet divergences and a pole due to infrared
divergences.  

Since the (finite-volume) tadpole sum and 
the (infinite-volume) integral share the same ultraviolet divergences,
it is convenient to introduce their difference~\cite{hasenleut}:
\begin{eqnarray}
&&\xi_s(L,M^2)=\frac{1}{L^3}\sum_{\vec\ell}\frac{1}{(\vec\ell^2+M^2)^s}\\
&&\qquad   -\frac{\sqrt{4\pi}\Gamma(s+1/2)}{\Gamma(s)}
       \int\frac{d^4q_E}{(2\pi)^4}\frac{1}{(q_E^2+M^2)^{s+1/2}}\,.\nonumber
\end{eqnarray}
$\xi$ can be evaluated as an integral of known mathematical 
functions~\cite{finvol}:
\begin{eqnarray}
\xi_s(L,M^2)&=&\frac{1}{(4\pi)^{3/2}\Gamma(s)}\\
&&\qquad  \times\int_0^\infty d\tau \tau^{s-5/2} e^{-\tau M^2}
    \left[\theta^3\left(\frac{L^3}{4\tau}\right)-1\right]\,.
\nonumber
\end{eqnarray}

In order to turn the previous NLO chiral expansions
into bare expansions, we can isolate 
the nonanalytic pieces in $\tilde{\sigma}_P$ 
and expand the rest in powers of quark masses:
\begin{eqnarray}
\frac{\tilde\sigma_P}{L^3}
 &=& \frac{\tilde{M}_P^2}{8\pi^2}
             \log\frac{\tilde{M}_P^2}{\mu^2}
     + \xi_{1/2}(L,\tilde{M}_P^2)\\
 &\to& \label{eq:redefsigma}
   \frac{{\rm LO}(\tilde{M}_P^2)}{8\pi^2}
             \log\frac{\tilde{M}_P^2}{\mu^2}
     +\frac{1}{L^3\tilde{M}_P}\\
 &&\qquad
        + \frac{{\rm LO}(\tilde{M}_P^2)}{8\pi^2}
             \log\frac{{\rm LO}(\tilde{M}_P^2)}{\tilde{M}_P^2}  
\nonumber\\ 
 &&     + \Bigg[\xi_{1/2}(L,{\rm LO}(\tilde{M}_P^2))
        -\frac{1}{L^3\sqrt{{\rm LO}(\tilde{M}_P^2)}}
        \Bigg]
 \nonumber\,,
\end{eqnarray}
where ${\rm LO}[M_P^2]$ denotes the leading-order contribution to
the meson mass:
\begin{eqnarray}
{\rm LO}[\tilde{M}_\pi^2]&=&qr Y(3)M_\pi^2\,,\\
{\rm LO}[\tilde{M}_K^2]&=&\frac{(q+1)r}{2} Y(3)M_\pi^2\,, \\
{\rm LO}[\tilde{M}_\eta^2]&=&\frac{(q+2)r}{3} Y(3)M_\pi^2\,.
\end{eqnarray}

In App.~\ref{app:sigmaP}, we check
that all the nonanalytic dependence on $\tilde{M}_P^2$ 
in $\tilde\sigma_P$ comes from the first two terms in
eq.~(\ref{eq:redefsigma}), whereas 
the last (bracketed) is analytic in ${\rm LO}(\tilde{M}_P^2)$. 
We could take eq.~(\ref{eq:redefsigma}) as the bare expansion
of $\tilde\sigma_P$. However, a further simplification
can be performed~: the (last) logarithmic term in
Eq.~(\ref{eq:redefsigma}) is small whatever the size of the
fluctuations and contributes to $\tilde\sigma_P$ only at NLO.
We choose to absorb these terms in the NNLO remainders of
the masses and decay constants and to define the
bare expansion of $\tilde\sigma_P$ through:
\begin{eqnarray}\label{eq:redefsigma2}
\frac{\tilde\sigma_P}{L^3}
&=&
  \frac{{\rm LO}(\tilde{M}_P^2)}{8\pi^2}\log\frac{\tilde{M}_P^2}{\mu^2}
    +\frac{1}{L^3\tilde{M}_P} \\
&&\qquad
    + \left[\xi_{1/2}(L,{\rm LO}(\tilde{M}_P^2))
              -\frac{1}{L^3\sqrt{{\rm LO}(\tilde{M}_P^2)}}\right]+\ldots
\nonumber
\end{eqnarray}
where the ellipsis denotes $O(p^4)$ terms that are
absorbed in the NNLO remainders.

First, let us notice that we recover the results of the previous section
in the large-volume limit: only the first term in
eq.~(\ref{eq:redefsigma2}) survives, which 
is exactly the bare expansion of the infinite-volume tadpoles
discussed in Sec.~\ref{sec:bareinf}.
In addition,
this choice for the bare expansion settles related issues concerning
the convergence of the expansion at finite volume.
$\chi$PT in a
finite box can be consistently formulated in three different regimes,
called $p$-, $\epsilon$- and $\delta$-expansions, depending
on the relative sizes of the box sides and the pion
mass~\cite{GLfinvol}. 
In particular, the $p$-expansion used here holds if:
\begin{equation} \label{eq:pregime}
F_\pi L\gg 1\,, \qquad \qquad M_\pi L \gg 1\,.
\end{equation}
If the second condition is violated, the pion is too large to
be contained in the box : the breakdown of the $p$-expansion is flagged
by infrared divergences.
In the case of small fluctuations, quark condensation is responsible
for $N_f=3$ chiral symmetry breaking. The second
condition in eq.~(\ref{eq:pregime}) is translated
into $2mB_0L^2={\rm LO}(M_\pi^2)\times L^2\gg 1$. 
If either $m$ or $B_0$ are too
small compared to the size of the box, the condition is violated and
infrared divergences occur in $\tilde\sigma_P$.

In the case of large fluctuations, the second condition in 
eq.~(\ref{eq:pregime}) has a different interpretation.
We do not assume any specific mechanism for $N_f=3$ chiral
symmetry breaking. For non vanishing quark masses,
$B_0$ may be small without leading to a small pion mass : for instance,
in the limit case $B_0\to 0$, a significant pion mass can be
generated by the vacuum fluctuations encoded in $L_6$ and
$L_4$. Therefore, a small pion mass is obtained only if the quarks are
light enough, and the $p$-expansion must break down for $m$ too small but
\emph{not} for $B_0$ too small. One can check
that the proposed bare expansion of $\tilde\sigma_P$ exhibits
an infrared divergence when $m\to 0$, but not $B_0\to 0$. In other
words, even a very small three-flavour quark 
condensate does not lead to the breakdown of the $p$-expansion if
vacuum fluctuations are significant enough.

We could now take the NLO finite-volume expressions
for the masses and decay constants, perform the replacement in 
eq.~(\ref{eq:redefsigma2}) and study the resulting expressions.
However, we are mainly interested in the finite-volume corrections
to the infinite-volume estimates:
\begin{equation}
\Delta(X)=\frac{\tilde{X}(L)-\tilde{X}(\infty)}{\tilde{X}(\infty)}\,.
\end{equation}

We obtain the following results for the decay constants:
\begin{eqnarray}
&&\Delta(F_\pi^2) =
      -\frac{1}{4\tilde{F}_\pi^2(\infty)}
         \left[4\tilde\Xi_\pi
            +2\tilde\Xi_K\right]\,,
\\
&&\Delta(F_K^2) =
     -\frac{1}{4\tilde{F}_K^2(\infty)}
           \left[\frac{3}{2}\tilde\Xi_\pi
                +3\tilde\Xi_K
                +\frac{3}{2}\tilde\Xi_\eta
           \right]\,,
\nonumber
\end{eqnarray}
where:
\begin{equation}
\tilde\Xi_P=
  \xi_{1/2}(L,{\rm LO}[\tilde{M}_P^2])
   -\frac{1}{L^3\sqrt{LO(\tilde{M_P^2})}}
   +\frac{1}{L^3\tilde{M_P}}\,.
\end{equation}

In a similar way, we obtain for the masses:
\begin{eqnarray} \label{eq:finvolfpimpi}
\Delta(F_\pi^2M_\pi^2) &=&
   -\frac{qrY(3)}{4\tilde{F}_\pi^2(\infty)}
         \frac{M_\pi^2}{\tilde{M}_\pi^2(\infty)}\\
&&\qquad \times
           \left[3\tilde\Xi_\pi
            +2\tilde\Xi_K
            +\frac{1}{3}\tilde\Xi_\eta
           \right]\nonumber\,,
\\
\Delta(F_K^2 M_K^2) &=&
   -\frac{(q+1)rY(3)}{8\tilde{F}_K^2(\infty)}
              \frac{M_\pi^2}{\tilde{M}_K^2(\infty)}\\
&&\qquad \times
           \left[\frac{3}{2}\tilde\Xi_\pi
            +3\tilde\Xi_K
            +\frac{5}{6}\tilde\Xi_\eta
           \right]\nonumber\,.
\end{eqnarray}

\subsection{Size of the finite-volume corrections}

\begin{figure*}
\begin{center}
\hbox{
\includegraphics[width=9cm]{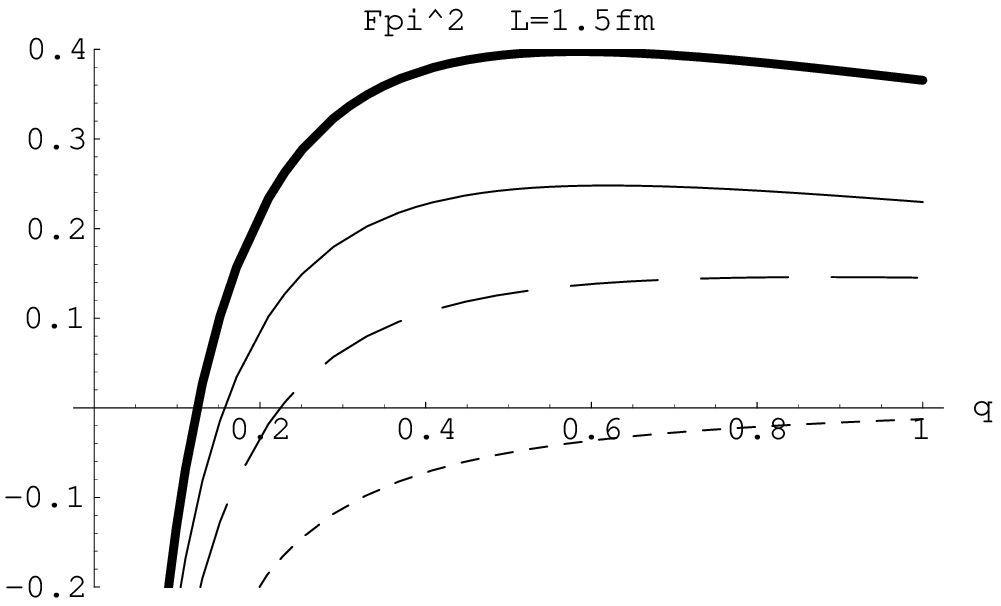}
\includegraphics[width=9cm]{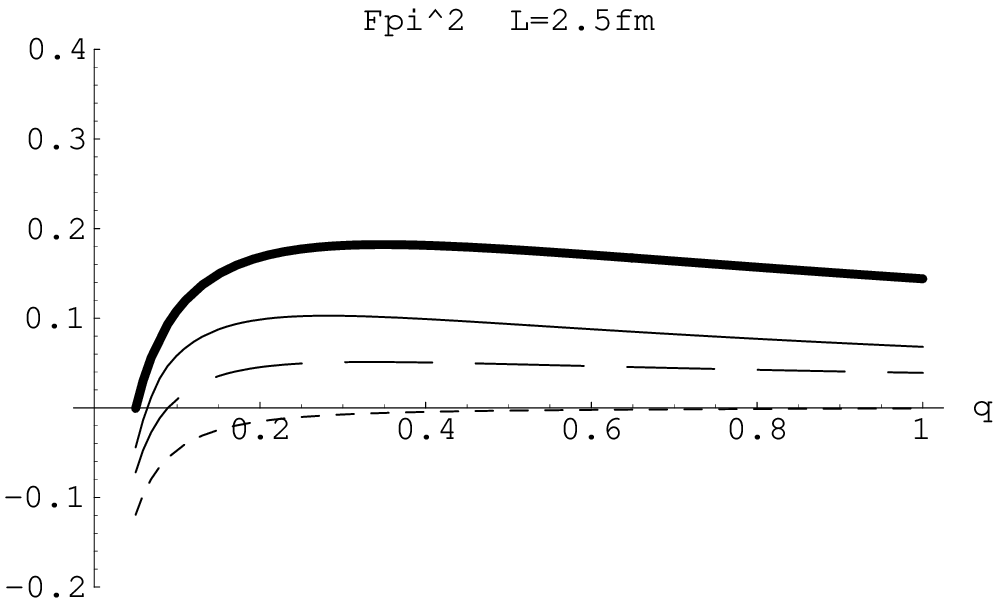}}
\hbox{
\includegraphics[width=9cm]{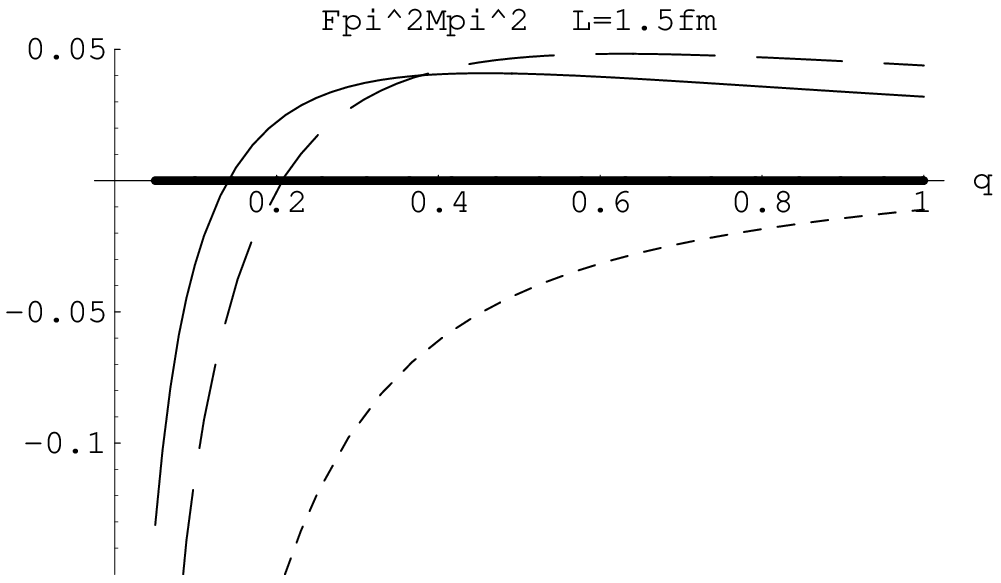}
\includegraphics[width=9cm]{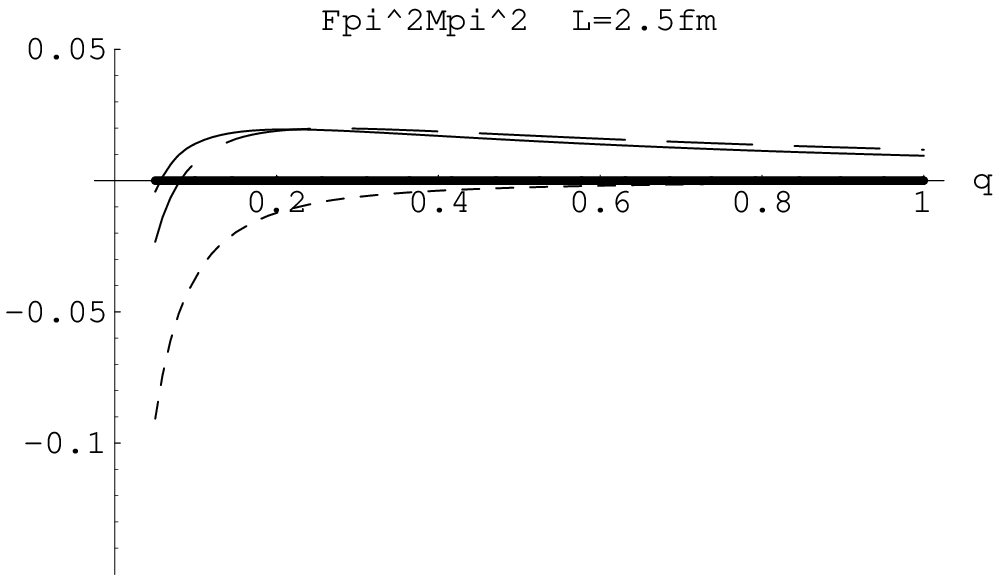}}
\end{center}
\caption{The relative finite-volume correction to $\tilde{F}_\pi^2$ (upper
plots) and $\tilde{F}_\pi^2\tilde{M}_\pi^2$ (lower plots). The left (right)
column corresponds to $L=1.5$ fm (2.5 fm). We set $r=25, Z(3)=0.8$. 
The thick full, full, long-dashed and dashed curves correspond to 
$X(3)=0,0.2,0.4,0.8$ respectively.}
\label{fig:samplefinvol}
\end{figure*}

We begin with a sample plot of the finite-volume corrections. 
In fig.~\ref{fig:samplefinvol},
$\Delta(F_\pi^2)$ and $\Delta(F_\pi^2 M_\pi^2)$ are plotted
as functions of $q$
for the specific choice of parameters $r=25$ and $Z(3)=0.8$. The left (right)
column corresponds to $L=1.5$ fm (2.5 fm), corresponding
to $M_\pi L=1.1$ ($M_\pi L=1.8$). The thick full, full, 
long-dashed and dashed curves correspond to $X(3)=0,0.2,0.4,0.8$
respectively. For $X(3)=0$, 
the finite-volume correction to $F_\pi^2M_\pi^2$ vanishes
at this order, see eq.~(\ref{eq:finvolfpimpi}).

We see that the corrections are very significant for $L=1.5$ fm, but
much smaller for $L=2.5$ fm. In addition the corrections are smaller 
for $F_\pi^2M_\pi^2$ than for the decay constant. Indeed, the 
finite-volume chiral expansion of $F_\pi^2M_\pi^2$ contain an
additional factor ${\rm LO}(\tilde{M}_\pi^2)$ multiplying the tadpole
terms $\tilde\sigma_P$. This tames the infrared 
divergences occurring when the pion
mass is too small compared to size of the box. This damping factor
is absent in the expansion of the decay constants.

\begin{figure*}
\begin{center}
\hbox{
\includegraphics[width=9cm]{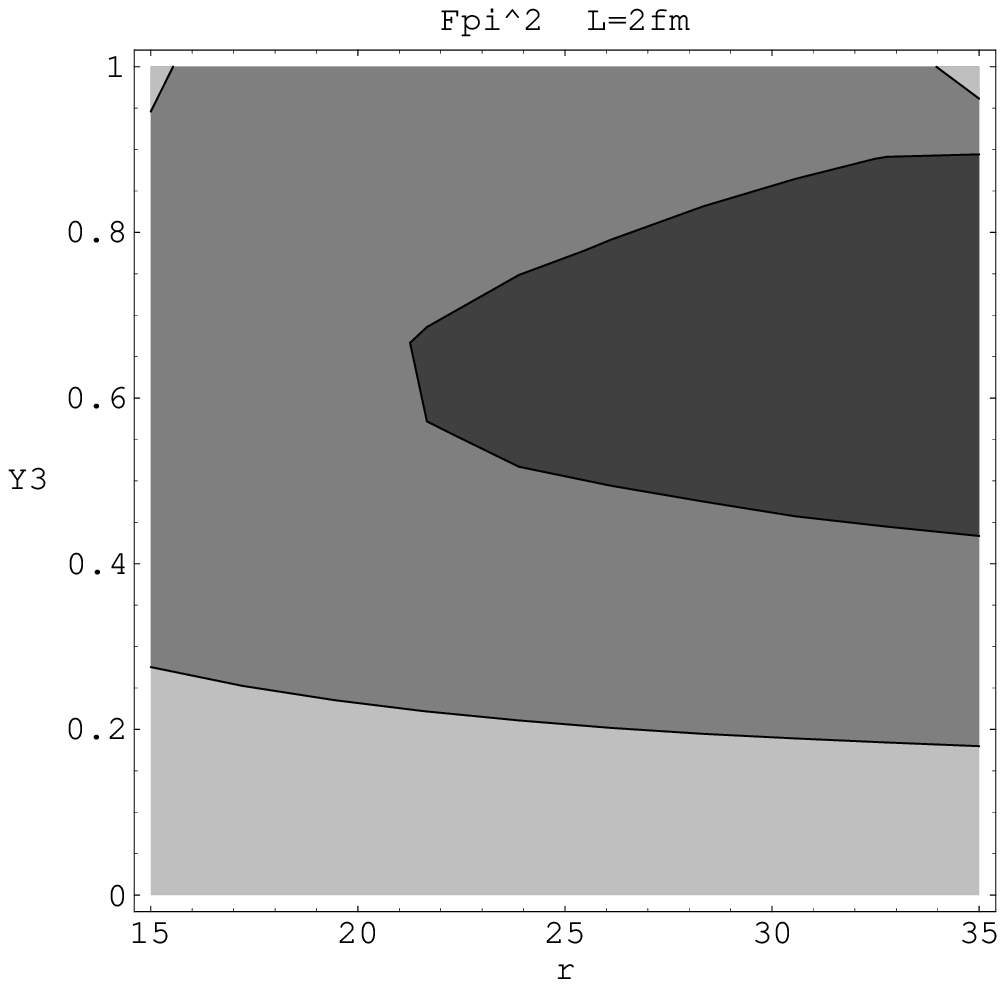}
\includegraphics[width=9cm]{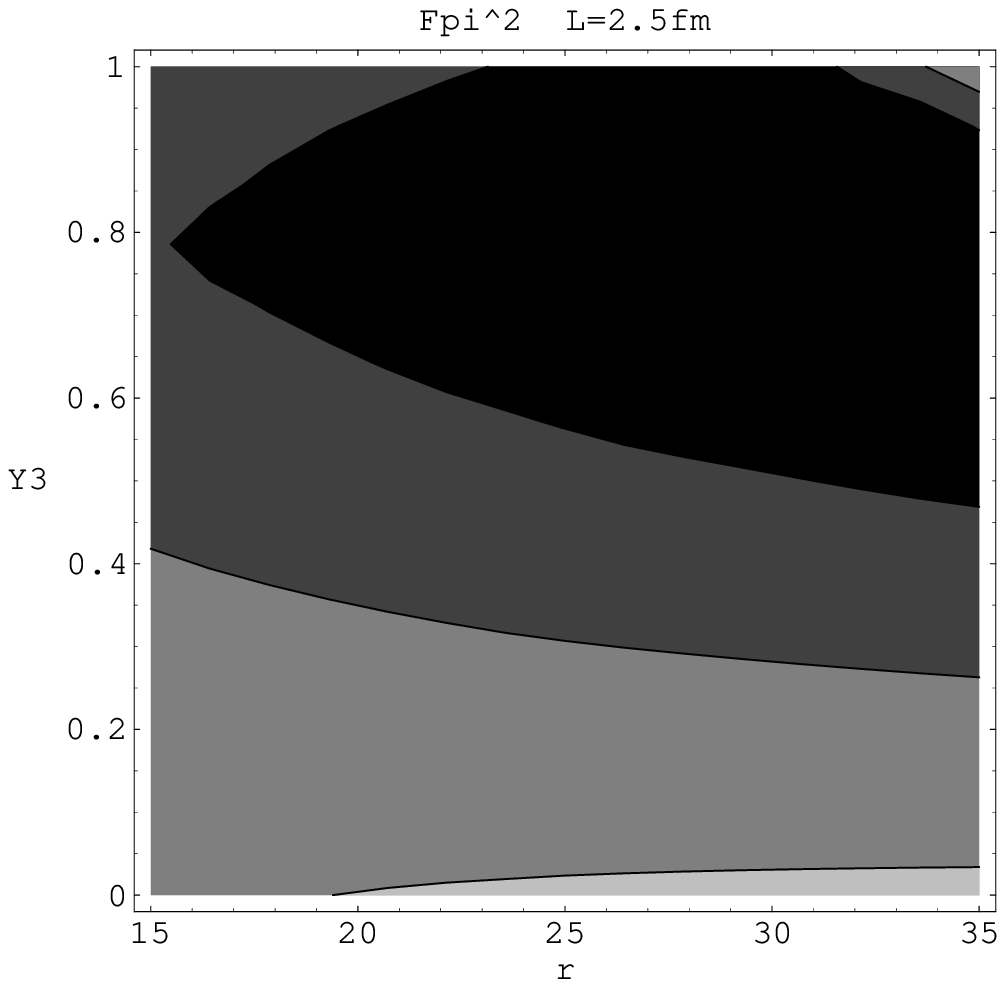}}
\hbox{
\includegraphics[width=9cm]{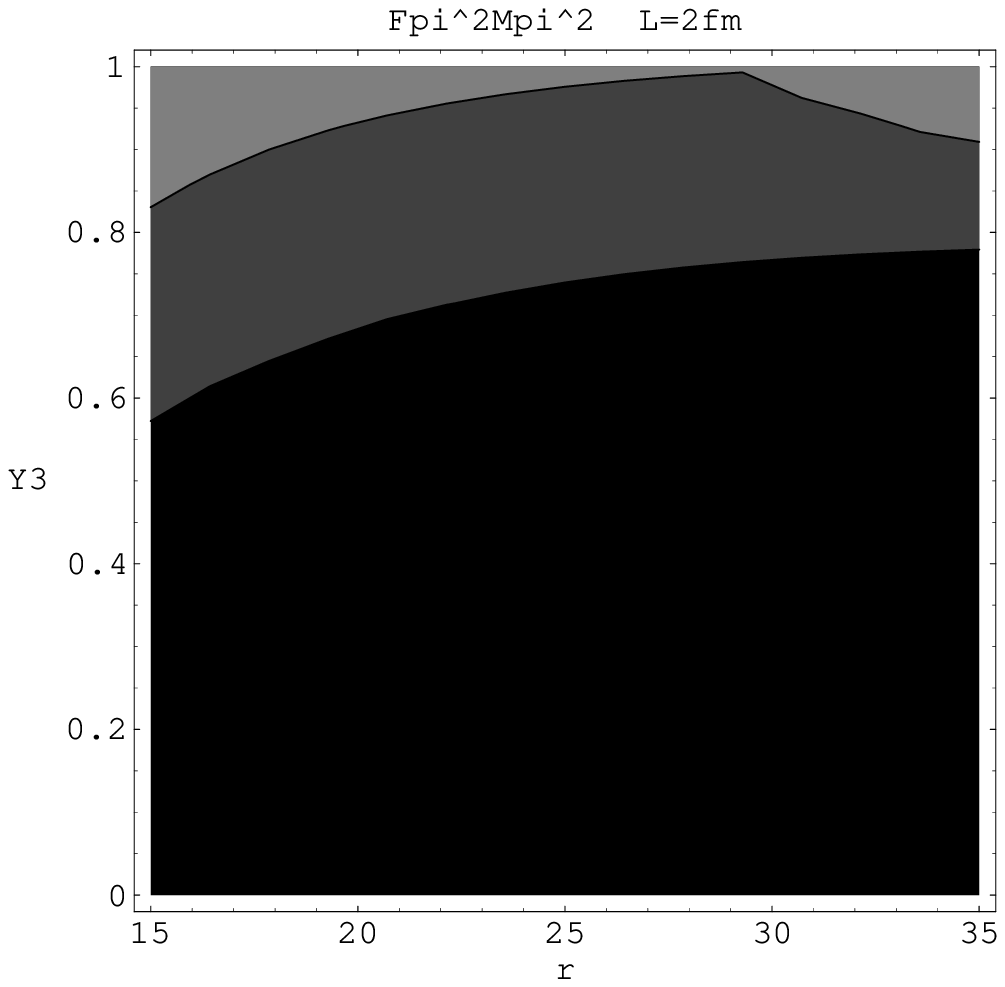}
\includegraphics[width=9cm]{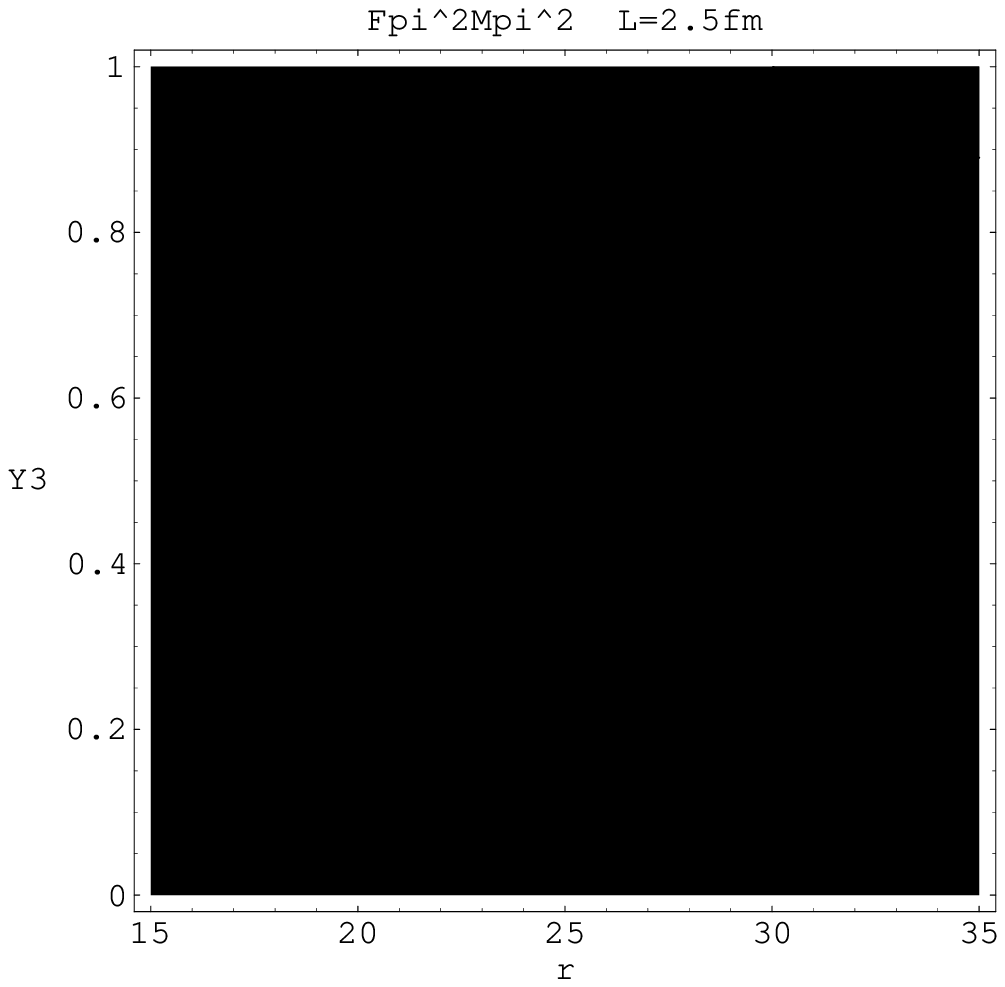}}
\end{center}
\caption{The maximal finite-volume corrections $D(\tilde{F}_\pi^2)$ 
(upper plots) and $D(\tilde{F}_\pi^2 \tilde{M}_\pi^2)$ (lower plots) 
as functions of $r$ and $Y(3)$. 
The left (right) column corresponds to $L=2$ fm (2.5 fm). 
Black domains correspond to $D$ smaller than 5\%,
increasingly lighter domains to $D$ smaller than 10,20 and 40\% 
respectively.}
\label{fig:pidelta}
\end{figure*}

\begin{figure*}
\begin{center}
\hbox{
\includegraphics[width=9cm]{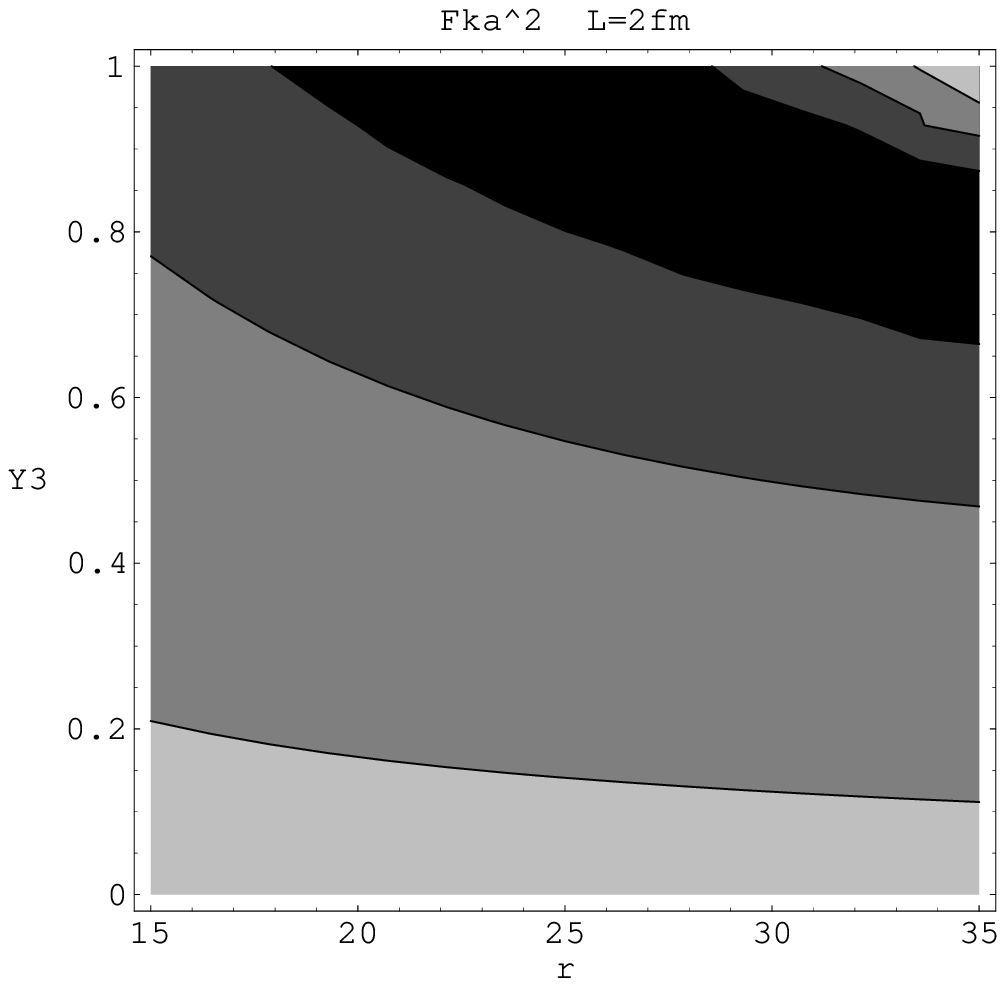}
\includegraphics[width=9cm]{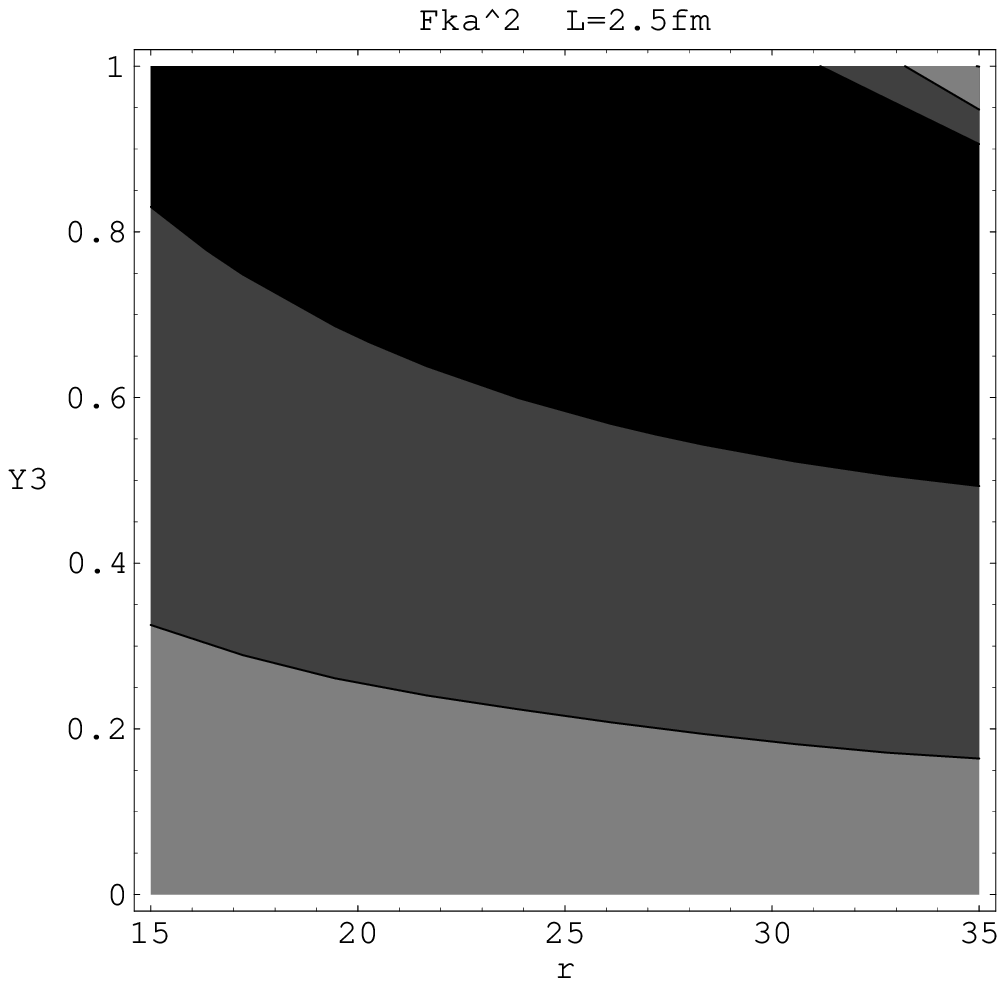}}
\hbox{
\includegraphics[width=9cm]{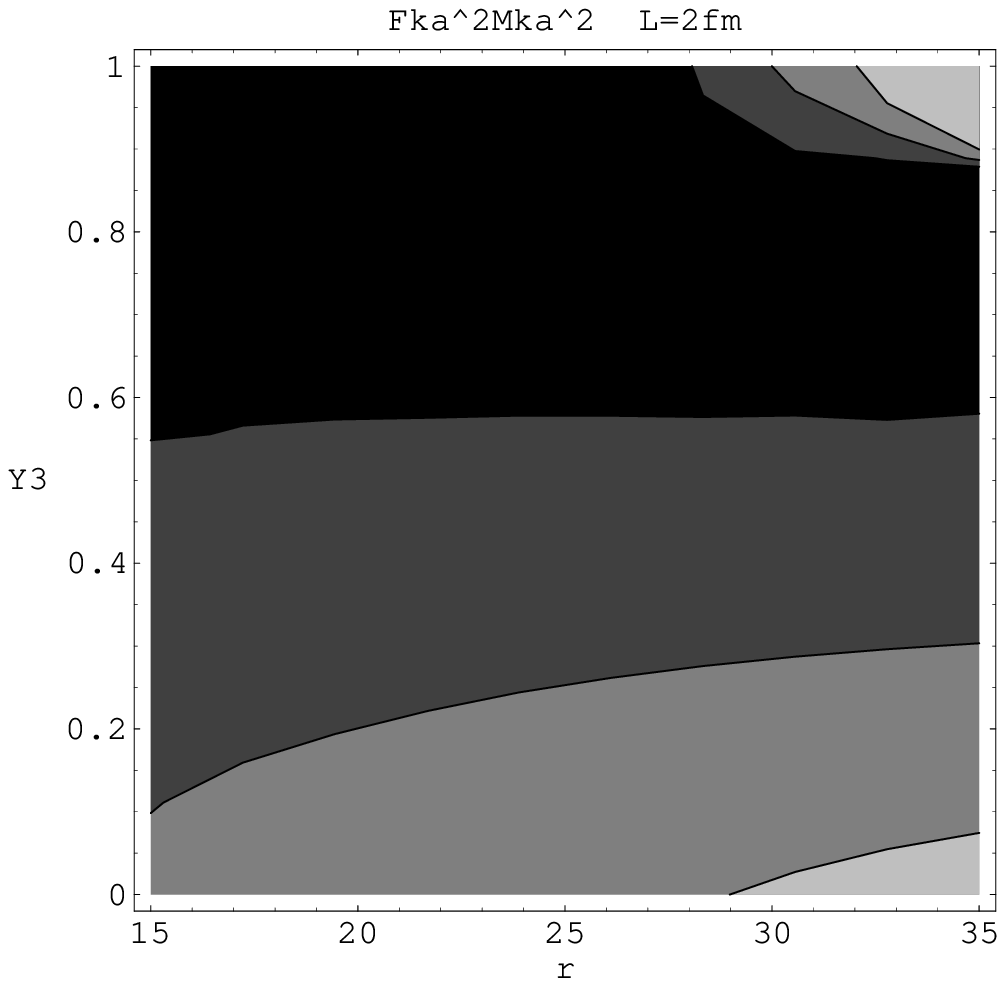}
\includegraphics[width=9cm]{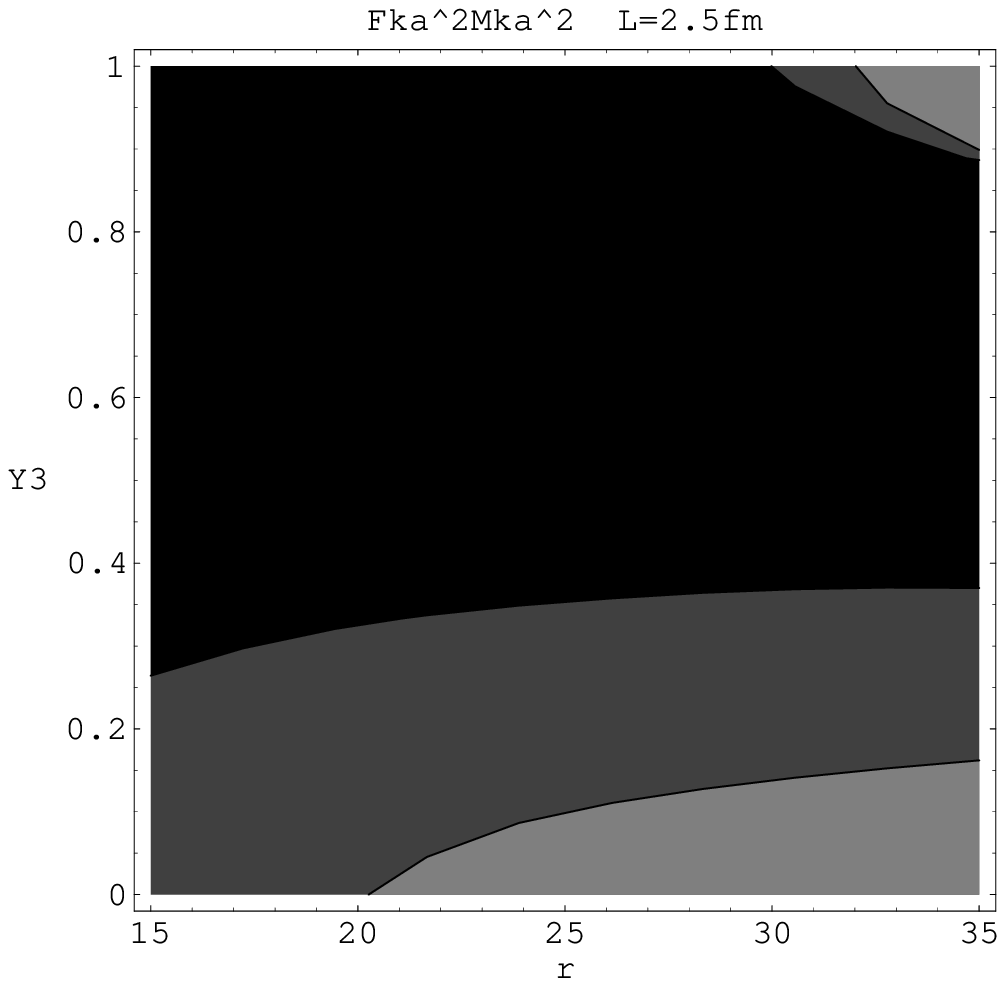}}
\end{center}
\caption{The maximal finite-volume corrections $D(\tilde{F}_K^2)$ (upper plots)
and $D(\tilde{F}_K^2 \tilde{M}_K^2)$ 
(lower plots) as functions of $r$ and $Y(3)$. 
The left (right) column corresponds to $L=2$ fm (2.5 fm). 
Black domains correspond to $D$ smaller than 5\%,
increasingly lighter domains to $D$ smaller than 10,20 and 40\% 
respectively.}
\label{fig:kadelta}
\end{figure*}

We want the finite-volume corrections to remain small over a reasonable
range of variation for $q$. Noticing that the size of the corrections
is weakly dependent on $Z(3)$, we introduce the quantity:
\begin{equation}
D(\tilde{X})= {\rm Max}\ |\Delta(\tilde{X})| \qquad
\left\{\begin{array}{c}
0.1\leq q \leq 1\\
0< Z(3) \leq 1
\end{array}\right.\,,
\end{equation}
whose size we investigate as a function of $r$ and $Y(3)$ for 
$L=2$ and 2.5 fm, corresponding respectively
to $M_\pi L=1.4$ and $M_\pi L=1.8$. In figs.~\ref{fig:pidelta} and \ref{fig:kadelta},
black regions correspond to $D$ smaller than 5\%,
increasingly lighter regions to $D$ smaller than 10,20 and 40\% 
respectively.

We observe the expected decrease of the finite-volume effects when
the size of the box increases. Once again, we notice that
the corrections to $F_P^2M_P^2$ are much smaller than those to the decay
constants, due to a better behaviour at the approach of the infrared 
region. $F_P^2M_P^2$ ($P=\pi,K$) in 
large volumes ($L=2.5$ fm) is a quantity for which
we manage a good control of finite-volume effects.

\section{Constraining three-flavour order parameters on the lattice}

For definiteness, we take a volume of size $L=2.5$ fm and set $r=25$.
However, we must keep in mind that the latter is a parameter 
which may vary between 15 and approximately 35. 
The qualitative conclusions that we
are about to draw do not depend on the exact values of $r$, but 
one observes small shifts in the following plots 
when $r$ is varied in its range.

We take into account the remainders associated with
the simulated masses and decay constants $\tilde{e}_P$ and 
$\tilde{d}_P$ (but not the indirect remainders $d,d',e,e'$).
As a rule of thumb~\cite{resum},
we estimate the size of NNLO remainders 
in the physical case by attributing a 30\% (10\%) effect
to an $m_s$ ($m$) factor, which leads to $O(m_s^2)\sim 10\%$,
$O(mm_s)\sim 3\%$ and $O(m^2)\sim 1\%$. 

For lattice simulations where $\tilde{m}$ varies, 
we take the following estimate:
\begin{eqnarray}
\tilde{d}_K &\sim& \tilde{e}_K = O(\tilde{m}m_s) \sim 0.10q^{1/3}\,,\\
\tilde{d}_K &\sim& \tilde{e}_\pi = O(m_s^2) \sim 0.10\,,
\end{eqnarray}
in order to recover the physical case, i.e.
 $O(\tilde{m}m_s)\simeq 0.10$ for $\tilde{m}=m_s$ ($q=1$)
and $O(\tilde{m}m_s)\simeq 0.03$ for $\tilde{m}=m$ ($q=1/r$).

\begin{figure*}
\begin{center}
\hbox{
\includegraphics[width=9cm]{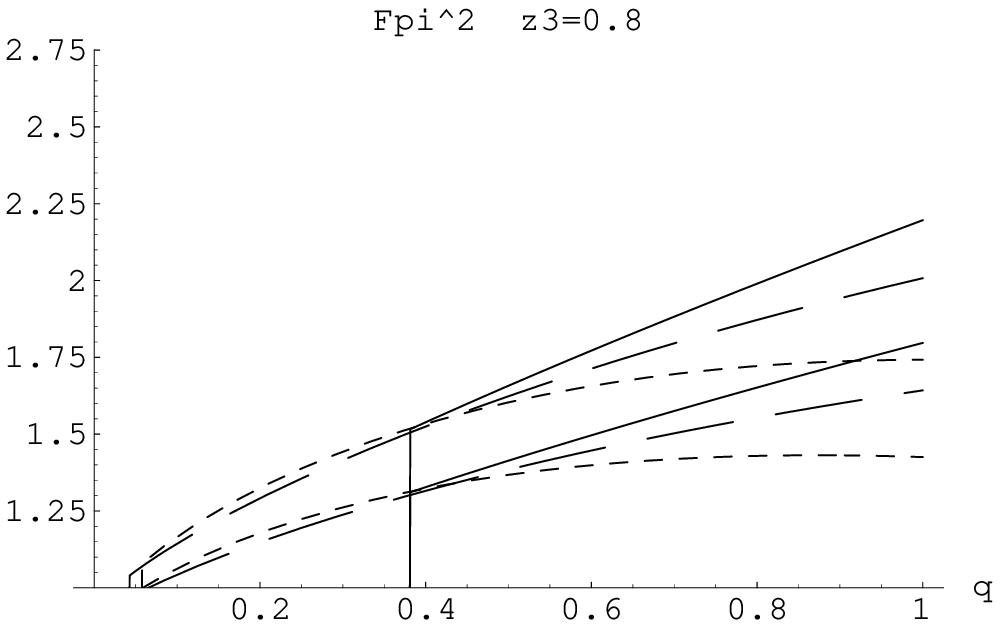}
\includegraphics[width=9cm]{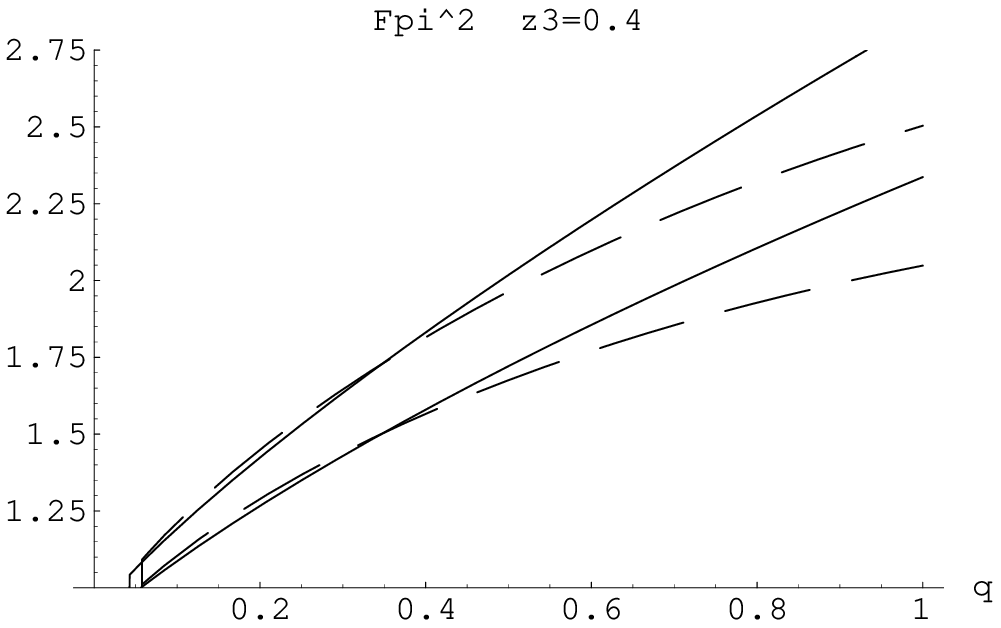}}
\hbox{
\includegraphics[width=9cm]{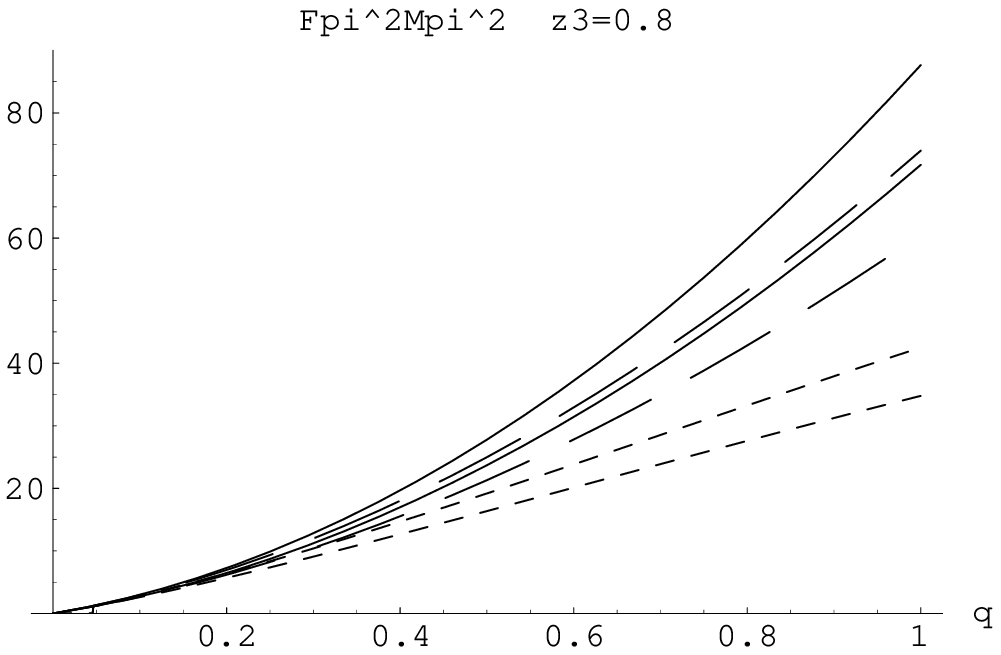}
\includegraphics[width=9cm]{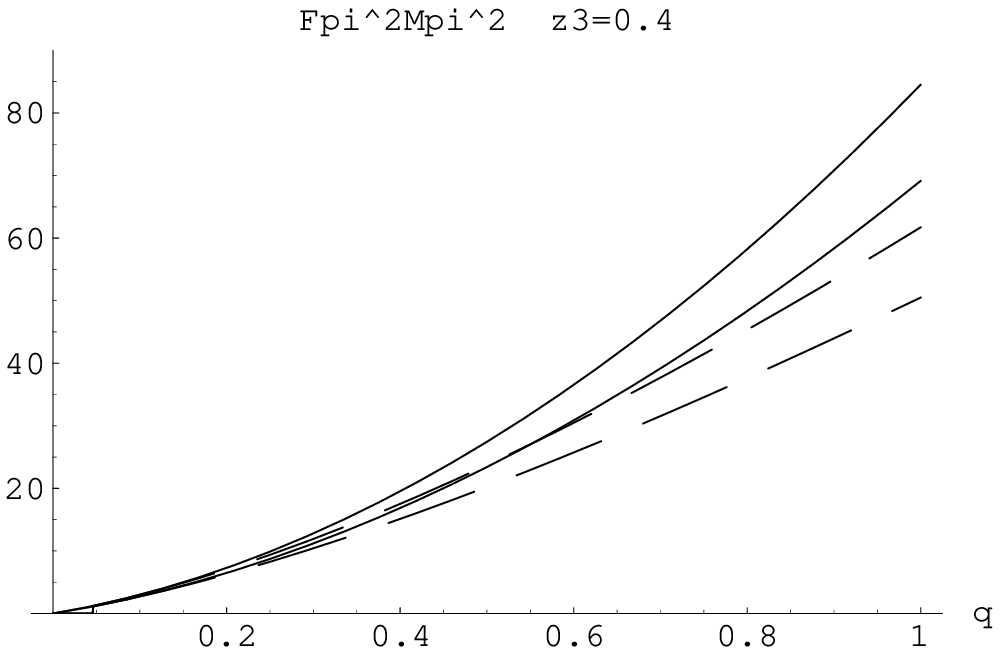}}
\end{center}
\caption{$\tilde{F}_\pi^2$ and $\tilde{F}_\pi^2\tilde{M}_\pi^2$ 
(normalised to their physical value) as
functions of $q$. The first (second) column corresponds to
$Z(3)=0.8$ (0.4). On each plot,
full, long-dashed and dashed bands correspond respectively
to $X(3)=0.8, 0.4, 0.2$. Cases where $Y(3)>1$ are not shown.}
\label{fig:plotuncpi}
\end{figure*}

\begin{figure*}
\begin{center}
\hbox{
\includegraphics[width=9cm]{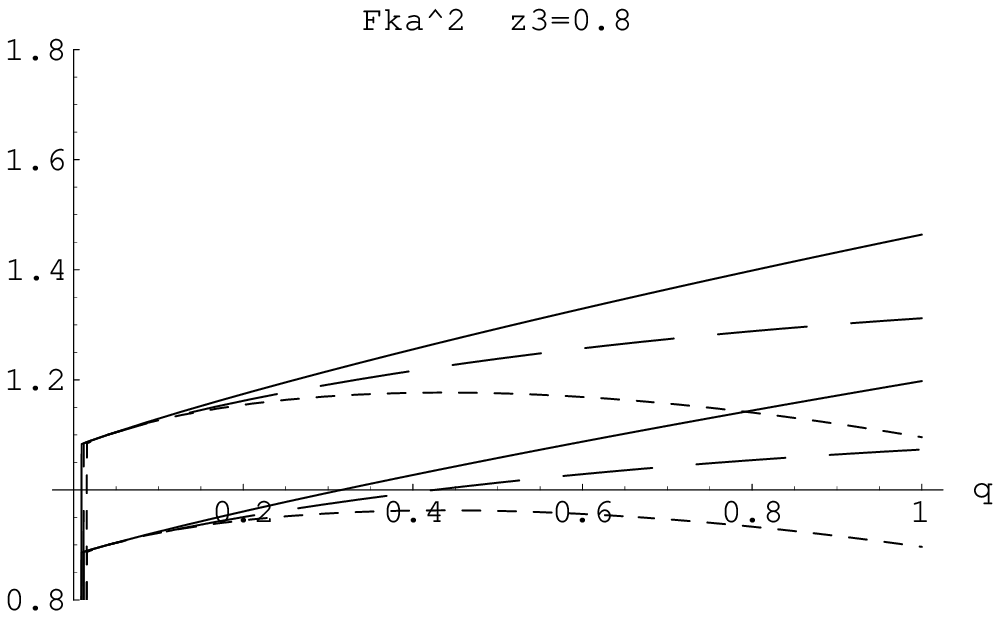}
\includegraphics[width=9cm]{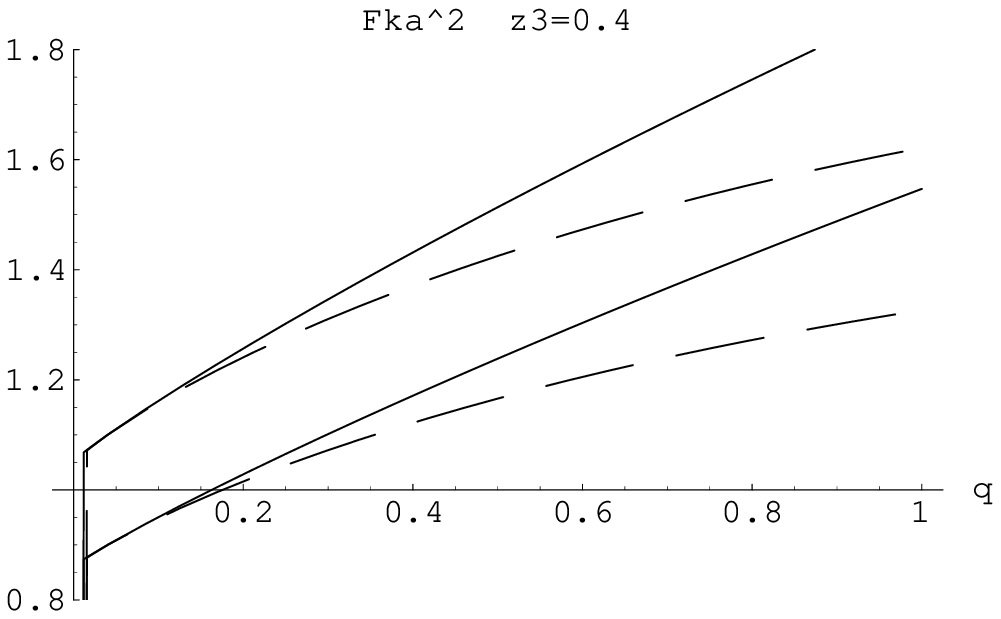}}
\hbox{
\includegraphics[width=9cm]{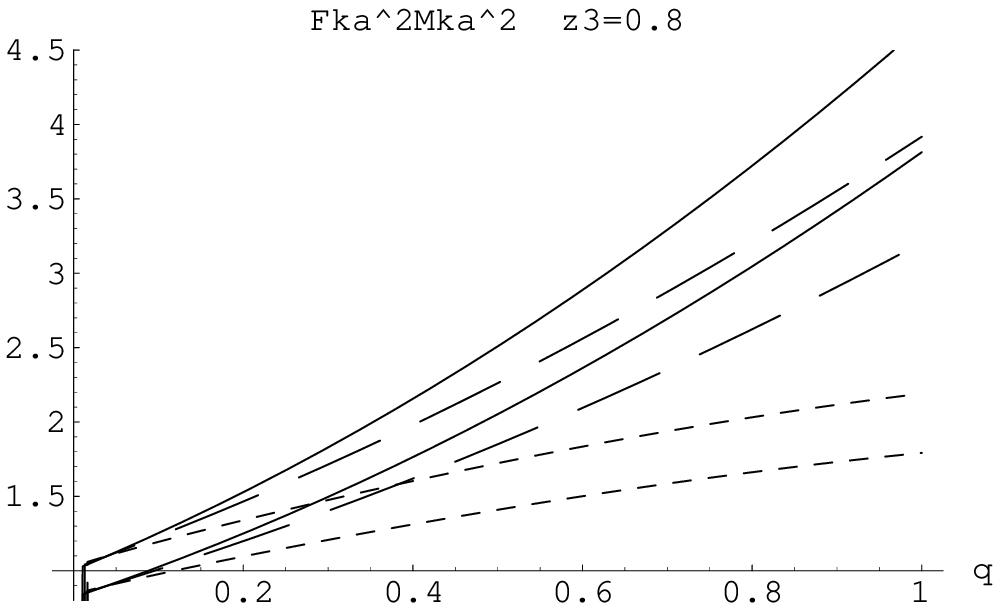}
\includegraphics[width=9cm]{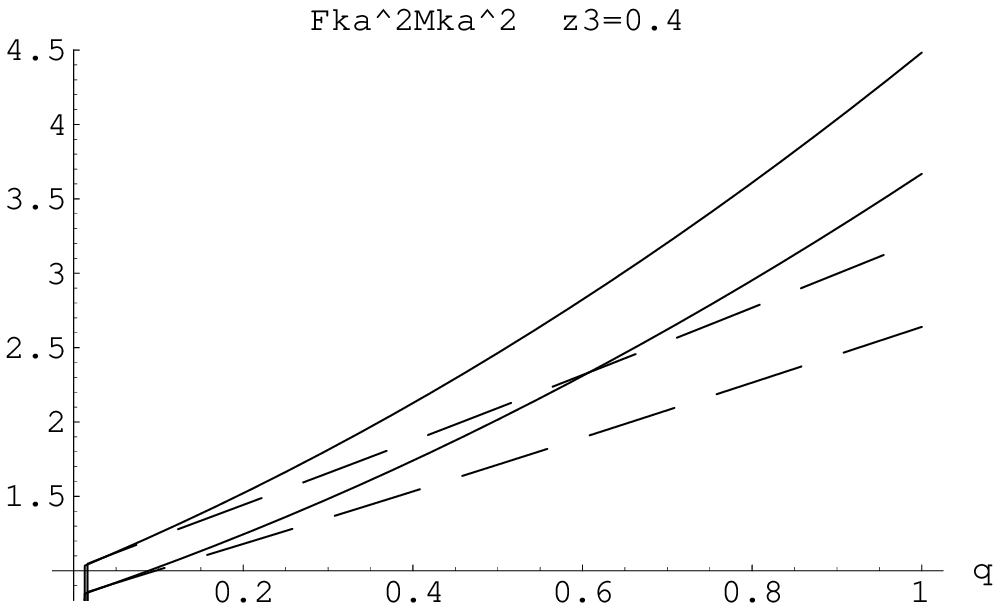}}
\end{center}
\caption{$\tilde{F}_K^2$ and $\tilde{F}_K^2\tilde{M}_K^2$ 
(normalised to their physical value) as
functions of $q$. The first (second) column corresponds to
$Z(3)=0.8$ (0.4). On each plot,
full, long-dashed and dashed bands correspond respectively
to $X(3)=0.8, 0.4, 0.2$. Cases where $Y(3)>1$ are not shown.}
\label{fig:plotuncka}
\end{figure*}

Figs.~\ref{fig:plotuncpi} and \ref{fig:plotuncka}
show the pseudoscalar decay constants and masses as
functions of the quark mass ratio $q=\tilde{m}/m_s$. 
The left column corresponds to $Z(3)=0.8$, the 
right one to $Z(3)=0.4$. On each plot, the various bands 
show the impact of NNLO uncertainties for 
different values of $X(3)$ [full: 0.8, long-dashed: 0.4, 
dashed: 0.2]. We plot the results only when 
the finite-volume effects are smaller than 10\%, but
we do not include these corrections in the quantities plotted.

At small $q$, the pion mass becomes too light, the finite-volume 
effects become larger than 10\%, and we cannot trust the corresponding
results because of potentially large higher-order corrections
(in such a case, we set the result to 0 on the plots). In particular the 
upper left plot in fig.~\ref{fig:plotuncpi}
[$\tilde{F}_\pi^2/F_\pi^2$ for $Z(3)=0.8$], where $q$ must be rather
large to tame finite-volume effects,
confirms that the chiral expansions of the decay 
constants may suffer from sizeable uncertainties at small 
simulated masses because of large finite-volume corrections.

From the previous analysis, the sensitivity to the three-flavour
order parameters through the curvature seems more important in
the case of the masses (which have smaller finite-volume corrections as well). 
We can isolate this effect by considering
the dimensionless ratios:
\begin{equation}
R_\pi=\frac{1}{q}
  \frac{\tilde{F}_\pi^2 \tilde{M}_\pi^2}{F_\pi^2 M_\pi^2}\,,
\qquad \qquad
R_K=\frac{2}{(q+1)}\frac{\tilde{F}_K^2 \tilde{M}_K^2}{F_K^2 M_K^2}\,.
\end{equation}

\begin{figure*}
\begin{center}
\hbox{
\includegraphics[width=9cm]{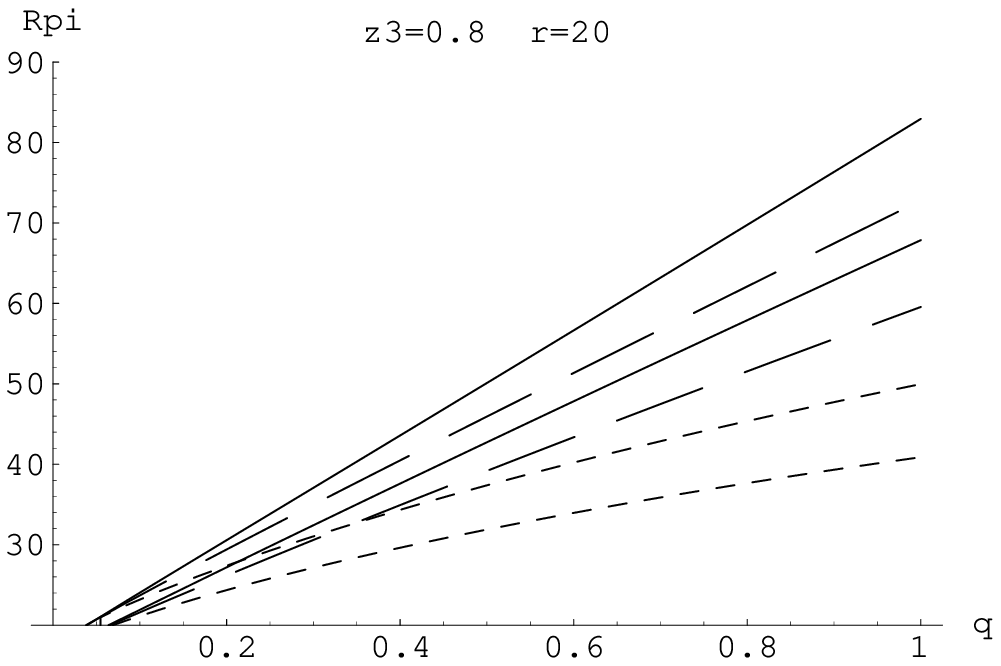}
\includegraphics[width=9cm]{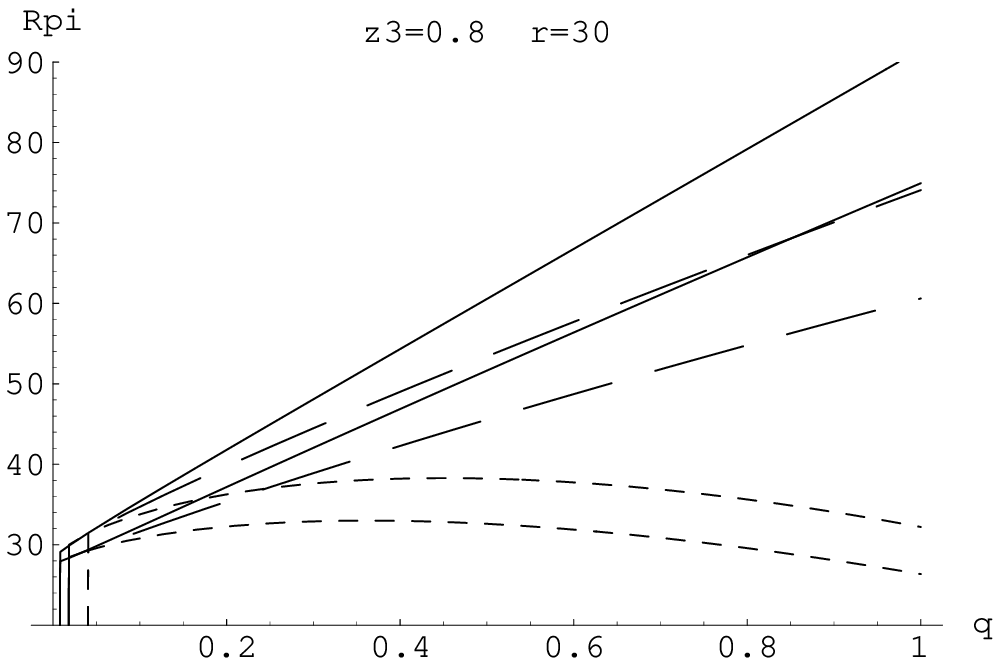}}
\hbox{
\includegraphics[width=9cm]{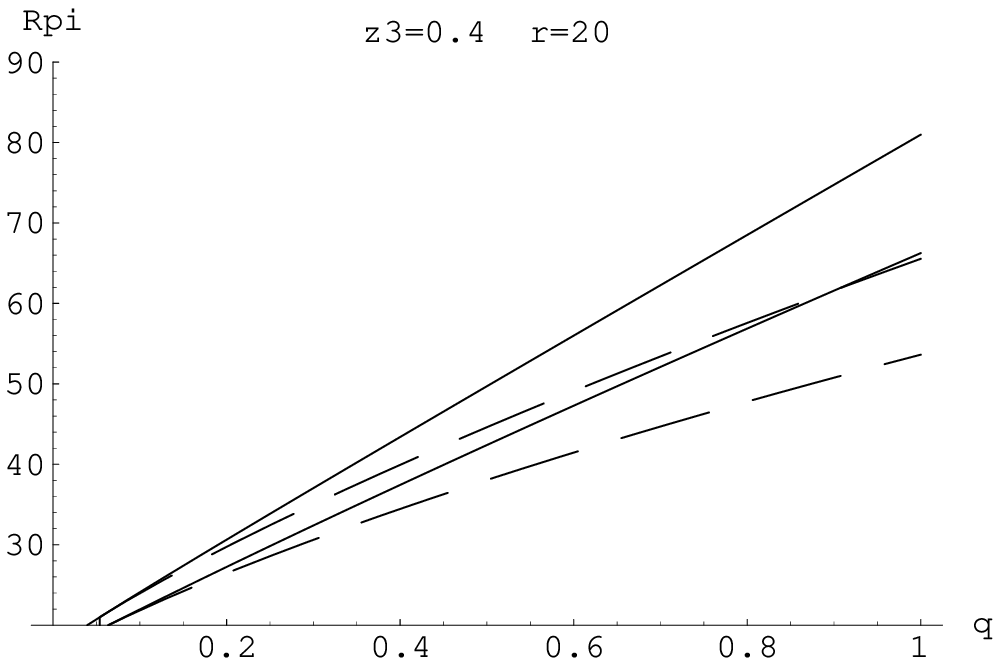}
\includegraphics[width=9cm]{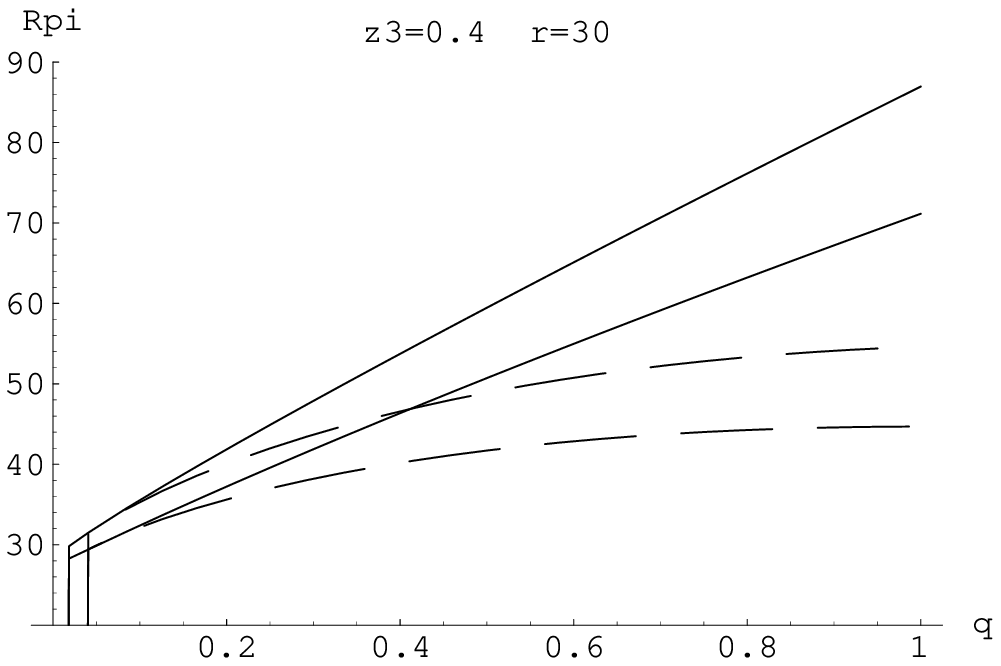}}
\end{center}
\caption{The ratio $R_\pi$ as a
  function of $q$. The first (second) column corresponds to
$r=20$ (30). The first (second) row deals with
$Z(3)=0.8$ (0.4). On each plot, 
full, long-dashed and dashed bands correspond respectively
to $X(3)=0.8, 0.4, 0.2$. Cases where $Y(3)>1$ are not shown.}
\label{fig:rpi}
\end{figure*}

\begin{figure*}
\begin{center}
\hbox{
\includegraphics[width=9cm]{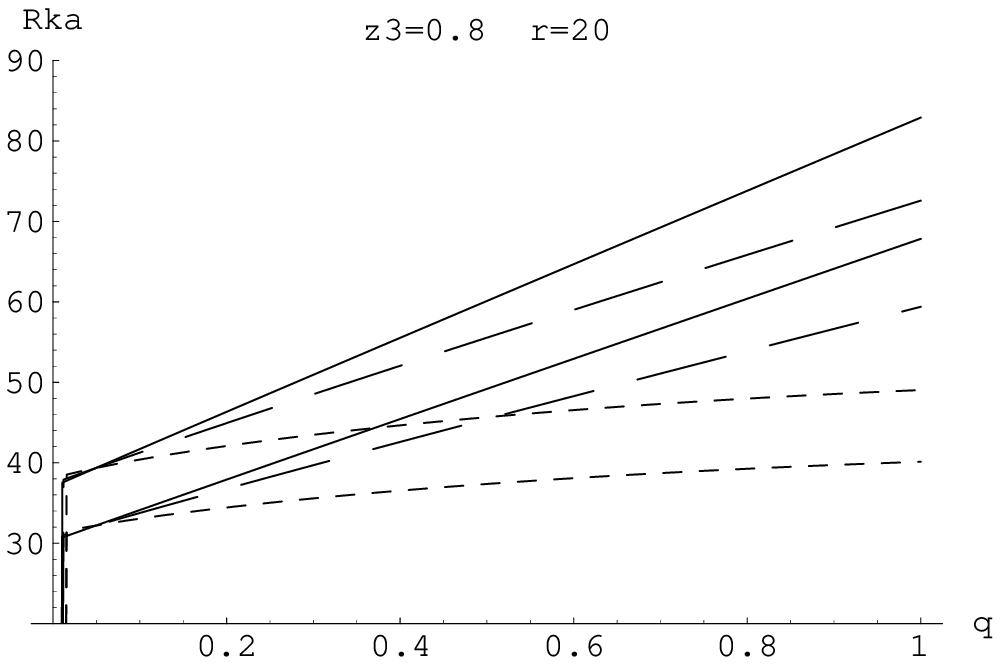}
\includegraphics[width=9cm]{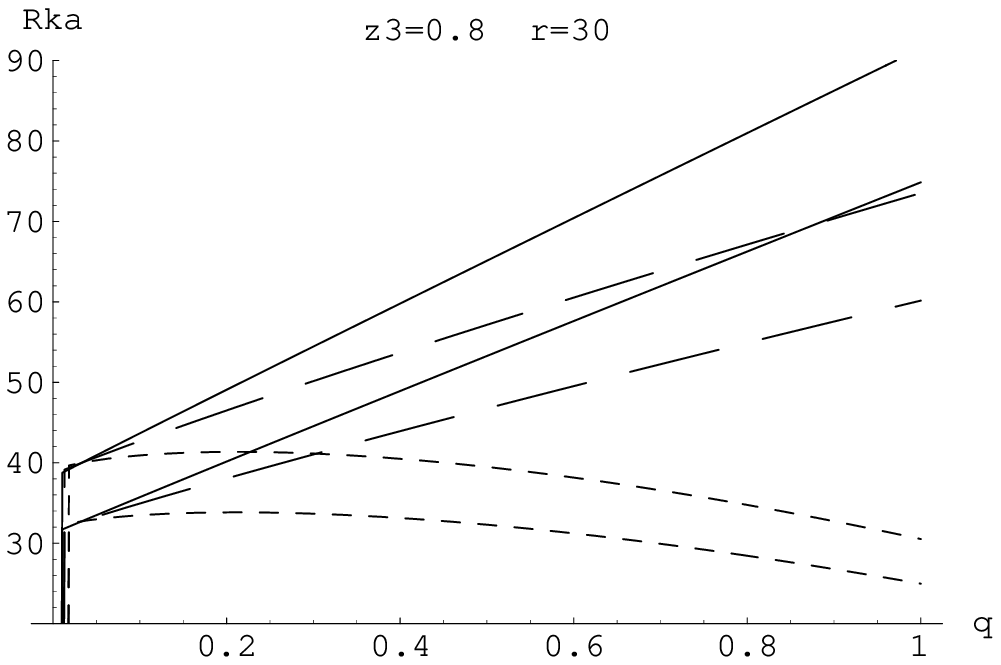}}
\hbox{
\includegraphics[width=9cm]{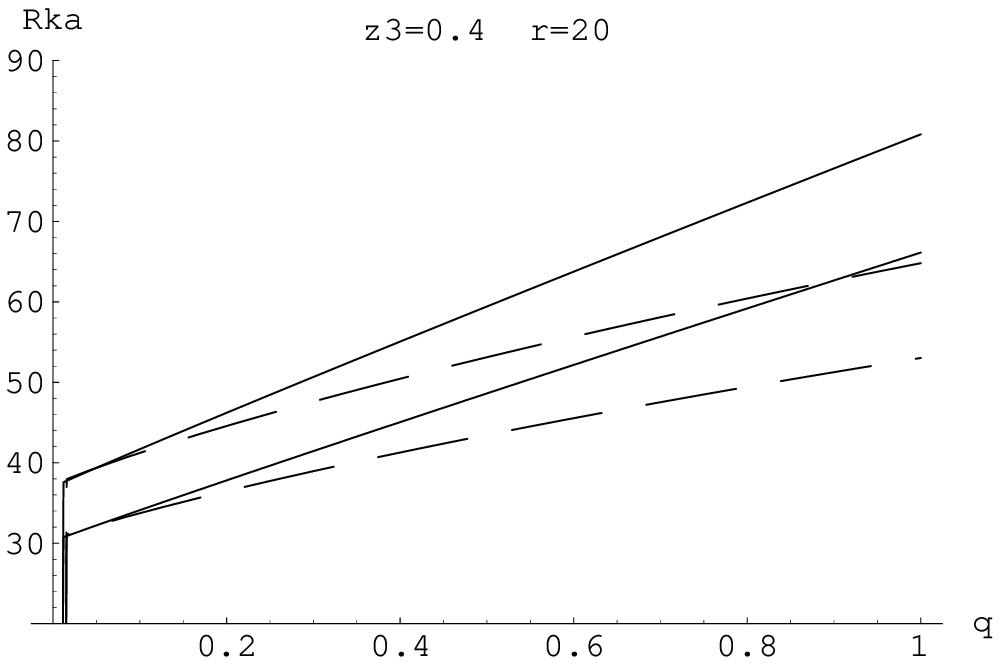}
\includegraphics[width=9cm]{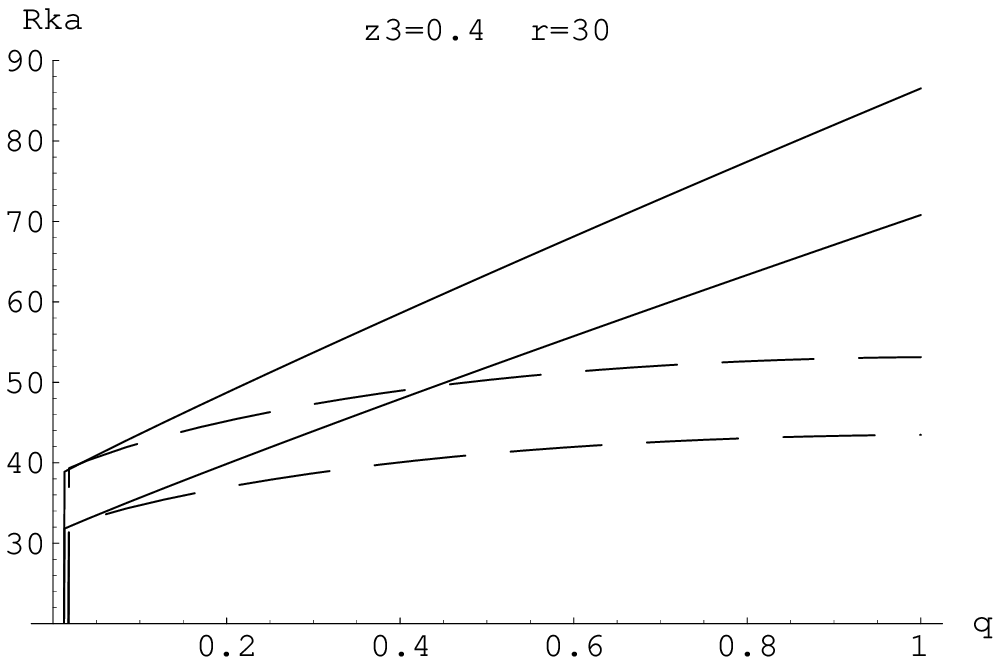}}
\end{center}
\caption{The ratio $R_K$ as a
  function of $q$. The first (second) column corresponds to
$r=20$ (30). The first (second) row deals with
$Z(3)=0.8$ (0.4). On each plot, 
full, long-dashed and dashed bands correspond respectively
to $X(3)=0.8, 0.4, 0.2$. Cases where $Y(3)>1$ are not shown.}
\label{fig:rka}
\end{figure*}

Figs.~\ref{fig:rpi} and \ref{fig:rka} indicate that
these two ratios can provide a way of constraining the size of
vacuum fluctuations by varying $q$ from $1/r$ to 1. The larger $r$, the
easier the distinction between small and large fluctuations, even though
the uncertainties due to NNLO remainders increase at the same time.

\section{Conclusion}

The presence of massive $s\bar{s}$-pairs in QCD vacuum may induce
significant differences in the pattern of chiral symmetry breaking
between $N_f=2$ and $N_f=3$ chiral limits, i.e., when
$m_s$ remains at its physical mass and when $m_s$ is set to 0. This
effect, related to the violation of the Zweig rule in the scalar
sector, may destabilise three-flavour chiral expansions numerically,
by damping leading-order (LO) terms proportional to $\Sigma(3)$ and $F^2(3)$,
and by enhancing next-to-leading-order (NNLO) terms containing the $O(p^4)$ 
Zweig-rule violating low-energy constants $L_4$ and $L_6$. In such case,
a more careful treatement of chiral expansions is required to avoid
uncontrolled corrections from higher chiral orders. 

In a previous work~\cite{resum}, we proposed
a consistent framework to take into account the possibility of large vacuum 
fluctuations. Indirect hints from 
dispersive estimates~\cite{uuss,uuss2,roypika} suggest that this effect might
be significant, but an experimental determination has not been 
achieved yet. In this paper, we proposed to probe the size of 
$s\bar{s}$-pairs fluctuations through lattice simulations with three
dynamical flavours, with a strange quark at its physical mass $m_s$
and two lighter flavours of mass $\tilde{m}$. We focused on
the masses and decay constants of the pions and kaons,
and worked out chiral expansions 
which should exhibit small NNLO remainders even when
LO and NLO terms compete numerically.
The dependence of
these observables on $\tilde{m}$ can provide useful constraints
on the structure of the $N_f=3$ chiral vacuum, and in particular
on the size of vacuum fluctuations. Conversely, this dependence
on vacuum fluctuations should stand as a warning about three-flavour chiral
extrapolations on the lattice, which might prove more delicate to
handle than usually assumed if vacuum fluctuations of
$s\bar{s}$ pairs are significant.

We have also estimated the corrections due to the
finite spatial dimensions used in the lattice simulation.
As expected, large volumes ($L\sim 2.5$ fm) are required to 
prevent these corrections from spoiling the predictivity of the chiral
expansions in the $p$-regime. 
These corrections do not seem enhanced in the case
of large vacuum fluctuations for the observables considered here.

Finally, we isolated two dimensionless ratios based on $F_\pi^2M_\pi^2$
and  $F_K^2M_K^2$
showing interesting features. They are not affected
very strongly by finite-volume corrections, and their dependence
on $\tilde{m}$ could provide interesting insights at the size of 
vacuum fluctuations. Therefore, it would be rather interesting to
perform a lattice study of these ratios with three dynamical flavours,
a reasonably large spatial box, and an action with good chiral
properties. The results might shed some light on the pattern of
$N_f=3$ chiral symmetry breaking and the low-energy dynamics of QCD.


\begin{acknowledgement}
I thank D. Becirevic for enjoyable discussions on finite-volume effects 
and for his critical reading of the first draft of this paper, 
L.~Girlanda,  C.~Haefeli and J.~Stern for comments on the manuscript.
Work partially supported by EU-RTN Contract EURIDICE (HPRN-CT2002-00311).
\end{acknowledgement}

\appendix

\section{Bare expansion of finite-volume tadpoles} \label{app:sigmaP}

The finite-volume tadpole 
$\tilde\sigma_P$ is related to its infinite-volume counterpart through
$\xi_{1/2}$ which can be reexpressed as:
\begin{equation}
\xi_{1/2}(L,M^2)
 = 2 \lim_{L_t \to\infty} g^{(4)}_{1}
 = \frac{1}{\sqrt{\pi}} g^{(3)}_{1/2}\,,
\end{equation}
where $g^{(d)}_r$ has been introduced in App.~A of ref.~\cite{hasenleut},
and the zero-temperature limit is given in eq.~(B.1) of
the same reference~\footnote{We take this opportunity
to correct a typo in this equation, which should read:
\begin{equation}
\lim_{L_d\to\infty} g^{(d)}_r = \frac{g^{(d-1)}_{r-1/2}}{\sqrt{4\pi}}
\end{equation}}. 

From App.~B of ref.~\cite{hasenleut}, it is
straightforward to determine the expansion of $g^{(3)}_{1/2}$ in 
powers of $M$:
\begin{eqnarray}
g^{(3)}_{1/2}&=&
  \frac{1}{\sqrt{4\pi}L^2}
   \Bigg[
     \frac{2\pi}{ML}-\frac{M^2L^2}{4\pi}\log(M^2L^2)\\
&&\qquad\qquad \nonumber
     +\sum_{n=0}^\infty \frac{\gamma_n}{n!} (ML)^{2n}
   \Bigg]\,,\\
\gamma_n
  &=& \left(-\frac{1}{4\pi}\right)^n
        \left[\alpha_{n+1/2}+\frac{3}{(2n+1)(n-1)}\right]
\,,\ n\neq 1\,,
\\
\gamma_1
  &=& -\frac{1}{4\pi} 
    \left[\alpha_{3/2}-\Gamma'(1)-\log(4\pi)-\frac{5}{3}\right]\,,
\end{eqnarray}
with $\alpha_{n+1/2}$ are known numerical coefficients (the first
few are displayed in table 1 of ref.~\cite{hasenleut}).

This leads to the following expansion of $\xi_{1/2}(L,M)$:
\begin{eqnarray}
\xi_{1/2}(L,M^2)&=&\frac{1}{ML^3}-\frac{M^2}{8\pi^2}\log(M^2L^2)\\
\nonumber
&&\qquad
                 +\frac{1}{2\pi}
     \sum_{n=0}^\infty \frac{\gamma_n}{n!} M^{2n} L^{2n-2}\,.
\end{eqnarray}
Thus, the nonanalytic dependence of $\tilde\sigma_P$
on ${\rm LO}(\tilde{M}_P^2)$ comes only from the pole and the logarithm
singled out in eq.~(\ref{eq:redefsigma}). 
According to our prescription, the bare expansion is obtained
once the polynomial terms are reexpanded in powers of quark masses.
At our order of accuracy, this amounts to replacing $\tilde{M}_P^2$
by ${\rm LO}(\tilde{M}_P^2)$ in the polynomial pieces, i.e.:
\begin{eqnarray}
\frac{\tilde\sigma_P}{L^3}
 &=& \frac{\tilde{M}_P^2}{8\pi^2}
             \log\frac{\tilde{M}_P^2}{\mu^2}
     + \xi_{1/2}(L,\tilde{M}_P^2)\\
\label{eq:sigmabare}
 &\to & \frac{{\rm LO}(\tilde{M}_P^2)}{8\pi^2}
             \log\frac{\tilde{M}_P^2}{\mu^2}\\
\nonumber
&&\qquad
     + \Bigg\{\frac{1}{\tilde{M}_P L^3}
     -\frac{{\rm LO}(\tilde{M}_P^2)}{8\pi^2}\log(\tilde{M}_P^2L^2)\\
&& \qquad
    +\frac{1}{2\pi}
     \sum_{n=0}^\infty \frac{\gamma_n}{n!} [{\rm LO}(\tilde{M}_P^2)]^{n}
 L^{2n-2}\Bigg\}
\nonumber\\
 &=& \frac{{\rm LO}(\tilde{M}_P^2)}{8\pi^2}
             \log\frac{\tilde{M}_P^2}{\mu^2}
     + \frac{1}{\tilde{M}_P L^3}\\
&&\qquad
     -\frac{{\rm LO}(\tilde{M}_P^2)}{8\pi^2}\log(\tilde{M}_P^2L^2)\\
&&\qquad
         +\frac{{\rm LO}(\tilde{M}_P^2)}{8\pi^2}
            \log[{\rm LO}(\tilde{M}_P^2)L^2]
\nonumber\\
&&\qquad
     + \Bigg[
       \xi_{1/2}(L,{\rm LO}(\tilde{M}_P^2))
         -\frac{1}{\sqrt{{\rm LO}(\tilde{M}_P L^3)}}\Bigg]
\nonumber\\
 &=& \frac{{\rm LO}(\tilde{M}_P^2)}{8\pi^2}
             \log\frac{\tilde{M}_P^2}{\mu^2}
     + \frac{1}{\tilde{M}_P L^3}\\
&&\qquad
         +\frac{{\rm LO}(\tilde{M}_P)}{8\pi^2}
                  \log\frac{{\rm LO}(\tilde{M}_P^2)}{\tilde{M}_P^2}
\nonumber\\
&&\qquad
     + \Bigg[
       \xi_{1/2}(L,{\rm LO}(\tilde{M}_P^2))
         -\frac{1}{\sqrt{{\rm LO}(\tilde{M}_P) L^3}}\Bigg]\,.
\nonumber
\end{eqnarray} 
The bracketed term has only a polynomial dependence on 
${\rm LO}(\tilde{M}_P^2)$. This expression yields 
eq.~(\ref{eq:redefsigma2}) and the bare expansion of $\tilde\sigma_P$.

\end{document}